\documentclass[fleqn,usenatbib]{mnras}

\usepackage{newtxtext,newtxmath}
\usepackage[T1]{fontenc}
\usepackage{ae,aecompl}
\usepackage{xspace}

\usepackage{graphicx}	% Including figure files
\usepackage{amsmath}	% Advanced maths commands
\usepackage{amssymb}	% Extra maths symbols

\usepackage{bm}
\let\vec\bm%

\usepackage{mleftright}
\let\left\mleft%
\let\right\mright%

\newcommand{\bs}[1]{\boldsymbol{#1}}

\newcommand{\be}{\begin{equation}}
\newcommand{\en}{\end{equation}}

\newcommand{\pder}[2]{\frac{\partial{#1}}{\partial{#2}}}
\newcommand{\pdder}[2]{\frac{\partial^2{#1}}{\partial{#2}^2}}
\newcommand{\der}[2]{\frac{d#1}{d#2}}

\newcommand{\f}{\frac}

\newcommand{\ex}{\vec{e}_x}
\newcommand{\ey}{\vec{e}_y}
\newcommand{\ez}{\vec{e}_z}

\newcommand{\ephi}{\vec{e}_\phi}
\newcommand{\er}{\vec{e}_r}

\newcommand{\para}{\parallel}
\newcommand{\kb}{k_{\textup {B}}}

\renewcommand{\b}{\bb{b}}
\newcommand{\dvx}{\delta \varv_{x}}
\newcommand{\dvz}{\delta \varv_{z}}

\newcommand{\dt}{\frac{\delta T}{T}}
\newcommand{\drho}{\frac{\delta \rho}{\rho}}
\newcommand{\dA}{\f{\delta A}{B}}
\newcommand{\mH}{m_{\textup{H}}}

\newcommand\bb[1]{\mbox{\boldmath{$#1$}}}

\newcommand{\va}{\varv_a}
\newcommand{\cs}{c}

\title[On the Kelvin-Helmholtz instability with smooth initial conditions]
{On the Kelvin-Helmholtz instability with smooth initial conditions --
Linear theory and simulations}

\author[Thomas Berlok and Christoph Pfrommer]{
Thomas Berlok\thanks{E-mail: tberlok@aip.de} and
Christoph Pfrommer
\\
Leibniz-Institut f{\"u}r Astrophysik Potsdam (AIP), An der Sternwarte 16, D-14482 Potsdam, Germany
}

\date{Accepted XXX. Received YYY; in original form ZZZ}

\pubyear{2018}

\usepackage{xcolor}

\begin{document}
\label{firstpage}
\pagerange{\pageref{firstpage}--\pageref{lastpage}}
\maketitle


\begin{abstract}
  The Kelvin-Helmholtz instability (KHI) is a standard test of
hydrodynamic and magnetohydrodynamic (MHD) simulation codes and finds
many applications in astrophysics. The classic linear theory considers
a discontinuity in density and velocity at the interface of two
fluids. However, for numerical simulations of the KHI such initial
conditions do not yield converged results even at the linear stage of
the instability. Instead, smooth profiles of velocity and density are
required for convergence. This renders analytical theory to be only
approximately valid and hinders quantitative comparisons between the
classical theory and simulations. In this paper we derive a linear
theory for the KHI with smooth profiles and illustrate code testing
with the MHD code \textsc{Athena}. We provide the linear solution for
the KHI with smooth initial conditions in three different limits:
inviscid hydrodynamics, ideal MHD and Braginskii-MHD. These linear
solutions are obtained numerically with the framework \textsc{psecas}
(Pseudo-Spectral Eigenvalue Calculator with an Automated Solver),
which generates and solves numerical eigenvalue problems using an
equation-parser and pseudo-spectral methods. The \textsc{Athena}
simulations are carried out on a periodic, Cartesian domain which is
useful for code testing purposes. Using \textsc{psecas} and analytic
theory, we outline the differences between this artificial numerical
setup and the KHI on an infinite Cartesian domain and the KHI in
cylindrical geometry. We discuss several astrophysical applications,
such as cold flows in galaxy formation and cold fronts in galaxy
cluster mergers. \textsc{psecas}, and the linear solutions used for
code testing, are publicly available and can be downloaded from the
web.
\end{abstract}


\section{Introduction}
The KHI is a hydrodynamic instability, which can arise when two
adjacent fluids have a relative velocity at their interface, or
alternatively, when there is velocity shear in a single continuous
fluid.  Despite its long history \citep{Helmholtz1868,Kelvin1871}, the
KHI is still a very active research topic due to its observed
prevalence in Nature.

On Earth,  the KHI is important in geophysics where it was originally
envisaged to be
responsible for ocean surface waves \citep{Kelvin1871} and cloud billows
\citep{von1890energie,Houze2014}, and is
believed to be responsible for mixing in the oceans \citep{Smyth2012}. In the
near-Earth environment, it is observed at the boundaries of coronal mass
ejections in the solar corona \citep{Foullon2011,Mostl2013}, at the
interface between the solar wind and the magnetosphere of the Earth
\citep{Dungey1963,Pu1983,Hasegawa2004}, and the magnetospheres of other solar
system planets such as Saturn and Jupiter \citep{Johnson2014}. Outside the
solar system, the KHI is theorized to be responsible for chemical mixing in
the Orion nebula \citep{Berne2012}, and it has been observed in relativistic
outflows
from active galactic nuclei \citep{Lobanov2001}.

In galaxy clusters, the KHI is observed to arise at the interface between a
galaxy and the intracluster medium (ICM), leading to
mixing of the stripped material \citep{Nulsen1982}.
The KHI is also observed in the ICM itself, where a sloshing cold front
can arise due to a minor merger of a galaxy cluster
\citep{Tittley2005,ZuHone2016}. The KHI is
theorized to be important for understanding whether cold streams feed galaxies
at high redshift
\citep{Mandelker2016,Padnos2018,Mandelker2018}
and it is seen to grow as
parasitic modes feeding off channel modes 
\citep{Pessah2009,Pessah2010,Murphy2015} in simulations of 
the magneto-rotational instability (MRI, \citealt{Balbus1991}). In the latter
scenario, the KHI is theorized to set 
the magnetic field strength in core-collapse supernovae by terminating the MRI
and preventing continued magnetic field amplification 
\citep{Rembiasz2016a}.

At a fundamental level, the KHI offers a possible transition to turbulence
from laminar shear flows as the billows can be subject to secondary
instabilities (e.g., \citealt{Smyth1991,Matsumoto2004,Thorpe2012}).
Capturing the physics of the resulting turbulent dissipation and mixing is
considered important for astrophysical computer simulations
\citep{Agertz2007}. The KHI has consequently become one of the standard
tests for astrophysical hydrodynamic and MHD codes.
In recent years, it was realized by the astrophysical community that a
smooth initial condition is required
in order for such numerical tests to converge as a
contact discontinuity remains unresolved when the numerical grid resolution is
increased  \citep{Robertson2010,McNally2012}. With this in mind,
\citet{McNally2012} used a smooth velocity profile to obtain converged
simulations of the linear regime of the KHI and \citet{Lecoanet2016} obtained
converged solutions of the nonlinear evolution of the KHI by also including
explicit dissipation in their simulations.
Smooth profiles are now routinely being used for test simulations (e.g.,
\citealt{Roediger2013,Ji2018}) but these simulations are
often compared with analytic theory that
made assumptions which are not fulfilled by the simulations. Common
assumptions for
the linear theory, that simplify the analysis, are on the boundary
conditions, that the flow is incompressible and that the
background density and velocity profiles are discontinuous.
Alleviation of all these assumptions, in particular the
discontinuous profiles, necessitates a numerical calculation of the growth rate
and the perturbations to the system.

For the hyperbolic tangent profile, such numerical investigations of the KHI
eigenmode structure have been performed for incompressible
hydrodynamics already by \citet{Michalke1964}, for compressible
hydrodynamics by \citet{Blumen1970,Hazel1972,Blumen1975,Drazin1977} and
for compressible, ideal MHD by \citet{Miura1982}.
As
their results were numerical in nature and require dedicated code to
reproduce, these results are however not as readily accessible as the
analytical expressions found in e.g., \citet{Chandrasekhar1961}, which seems to
explain the widespread comparison with the theory found therein.
Three notable exceptions are the simulations by \citet{Miura1984},
\citet{Ryu1995} and  \citet{Frank1996}
who used the results
of \citet{Miura1982} for studying the MHD version of the KHI on a truncated,
infinite domain.
Linear theory indeed often assumes as boundary condition that the KHI
disturbances go to zero (infinitely) far away from the shear interface. For
many astrophysical codes, this is not an optimal boundary condition, and a
periodic  domain is therefore often considered instead.

In order to facilitate future code tests with an apples-to-apples comparison
between linear theory and simulations, it therefore seems prudent to
\emph{i)} update the linear theory such that it matches common simulation
setups and \emph{ii)} make the obtained linear solutions easily accessible to
the community. This is the main aim of the present paper in which we provide
the linear theory for a smooth
background profile with the doubly periodic boundary conditions that are
commonly employed in code tests. The starting point is
the hydrodynamic profile used by \citet{Lecoanet2016} but we include
additional physics such as a background magnetic field and anisotropic
viscosity which are important for astrophysical applications. In addition, we
discuss how the behavior of this system changes at high flow speeds where
surface
modes are replaced by body modes and the incompressible assumption is
inadequate.
Our linear solutions are obtained using a pseudo-spectral
method with the new Python package \textsc{psecas}
(\textbf{P}seudo-\textbf{S}pectral \textbf{E}igenvalue \textbf{C}alculator
with an
\textbf{A}utomated \textbf{S}olver) that we make freely available in an
effort to make such linear calculations more widely accessible.

\begin{figure}
\includegraphics[trim= 0 5 0 0]{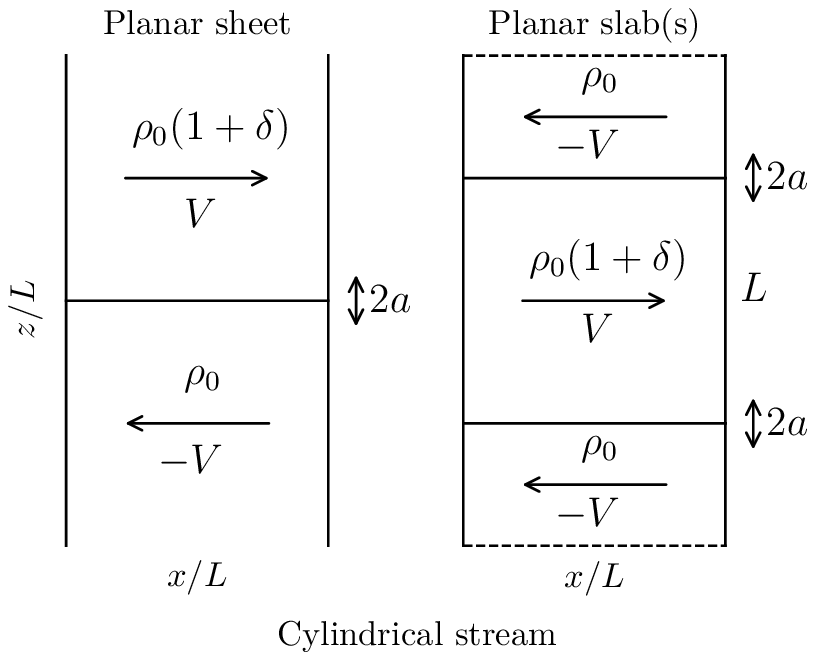}
\centering
\includegraphics[width=0.7\columnwidth, trim= 0 0 0 0, clip]{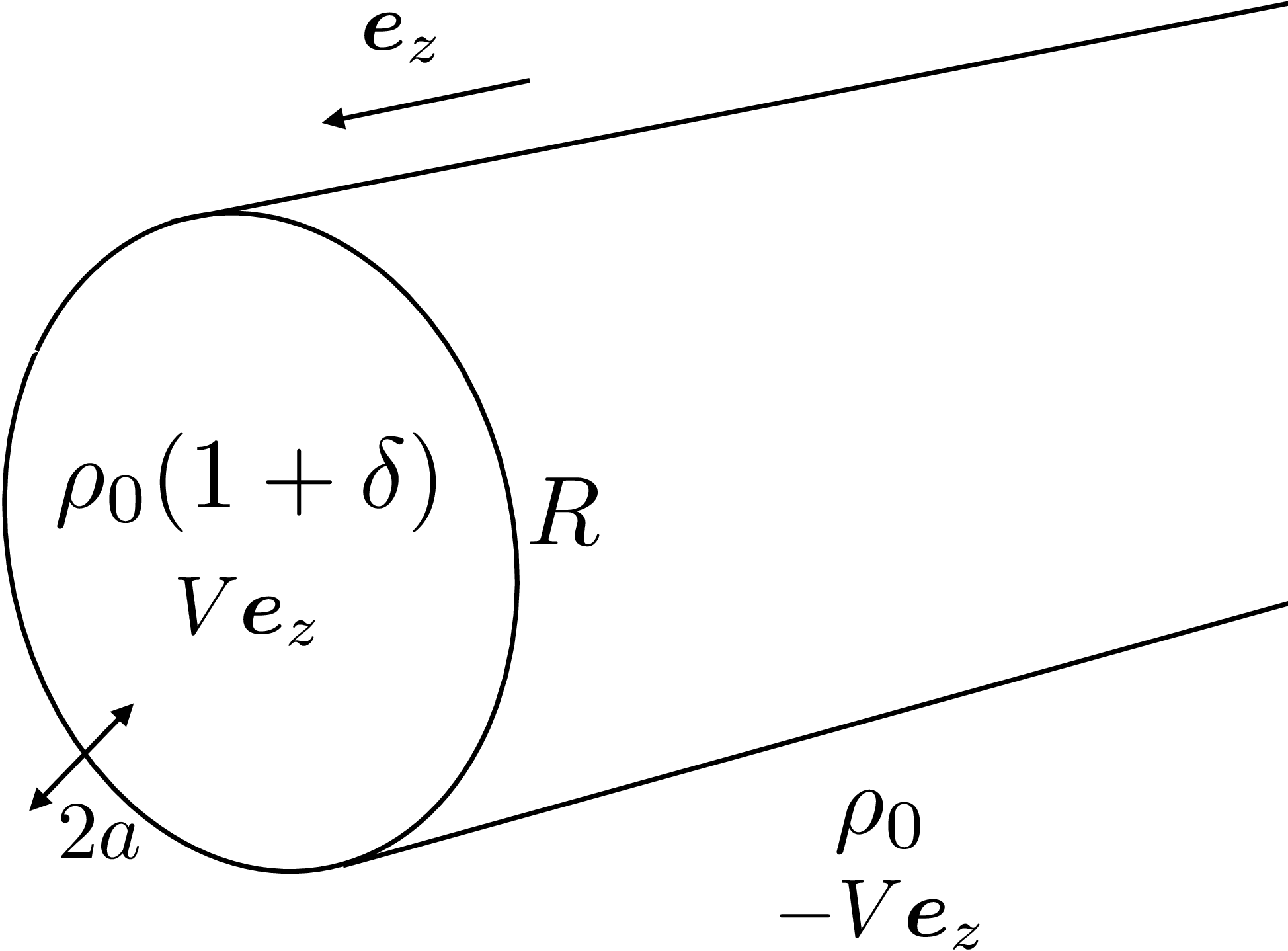}
\caption{The planar sheet with one interface where the velocity changes sign
(left), the planar slab (right) where the velocity changes sign twice and the
cylindrical stream (bottom) where the velocity changes sign at
$r=R$. We study the planar slab with two different boundary conditions in $z$:
\emph{i)} an infinite domain where disturbances are zero far from the shear
interfaces and \emph{ii)} a periodic domain which consists of two connected
slabs where we identify the dashed boundaries.
}
\label{fig:khi_setups}
\end{figure}

The remainder of the paper is divided as follows:
in Section~\ref{sec:khi-intro} we introduce the KHI as it arises in the planar sheet,
the planar slab and for the cylindrical stream, see Fig.~\ref{fig:khi_setups}.
Using the planar sheet as the simplest example, we explain the problem with
discontinuous profiles and how smooth profiles resolve this problem.
In Section~\ref{sec:equations_and_equilibria} we introduce the governing equations
along with their linearized form in Section~\ref{sec:linearized-equations}.
The linearized equations constitute an eigenvalue problem, which we solve using
the new \textsc{psecas} package, which we introduce in some detail in
Section~\ref{sec:psecas}.
In Section~\ref{sec:simulations}, we use the obtained linear solutions for the KHI
with smooth initial conditions on a periodic Cartesian domain and
illustrate how they can be used for quantitative comparisons with simulations.
We then present additional linear calculations of KHI eigenmodes in
Section~\ref{sec:bodymodes} in order to highlight qualitative differences
between the KHI in the various geometries. We discuss two astrophysical 
applications of the linear theory in Section~\ref{sec:discussion} and conclude 
by summarizing our results in Section~\ref{sec:conclusions}.

\section{The Kelvin-Helmholtz Instability}
\label{sec:khi-intro}

\begin{figure}
\includegraphics[trim= 0 25 0 0]{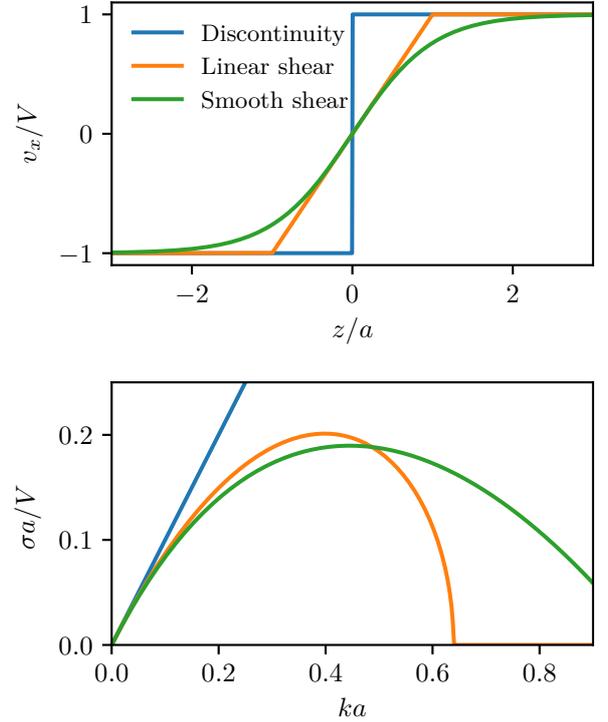}
\caption{Introducing a smooth change in the direction of the velocity on a
scale $\sim 2a$ prevent the KHI with wavelengths $\lambda \lesssim 2\pi a$
from growing. With sufficient spatial resolution, a smooth profile thus makes
it possible to resolve the full
spectrum of the linear instability in numerical simulations.}
\label{fig:smooth_growth_illu}
\end{figure}

The simplest system in which the KHI can take place is the planar sheet, see
Fig.~\ref{fig:khi_setups}. For this configuration, the flow velocity in the
$x$-direction changes sign at $z=0$ with $\varv_x=-V$ for $z<0$ and
$\varv_x=V$ for $z>0$. If we also allow for a difference in density with
$\rho=\rho_0$ at the bottom and $\rho=\rho_0(1 + \delta)$ at the top (where
$\delta > -1$) and assume that disturbances are zero as $z\rightarrow\pm
\infty$, then the dispersion relation for the KHI becomes (e.g.,
\citealt{Chandrasekhar1961}) \be \omega_\pm = \f{\delta \pm i
  2\sqrt{1+\delta}}{2 + \delta} k V \ ,
    \label{eq:chandra-khi}
\en
in the incompressible, inviscid hydrodynamic limit\footnote{The
dispersion relation for the planar sheet with compressible hydrodynamics and a
detailed analysis of it is presented in \citet{Mandelker2016}.}. Here $k$
is the wavenumber and the growth rate, $\sigma$, is given by the imaginary
part of $\omega$, i.e., $\sigma \equiv -\mathrm{Im}(\omega)$.
The discontinuous velocity profile is shown in the upper panel
of Fig.~\ref{fig:smooth_growth_illu} and the resulting growth rate with
$\delta=0$ is shown in the lower panel of Fig.~\ref{fig:smooth_growth_illu} with
blue solid lines.
Equation~\eqref{eq:chandra-khi} and Fig.~\ref{fig:smooth_growth_illu} show that the growth
rate is proportional to the wavenumber such that doubling the grid resolution
leads to a doubling of the growth rate of the fastest growing, resolved mode.
This explains why a computer simulation will never converge with a
discontinuous profile.
In
practice, numerical dissipation will damp wavenumbers close to the grid scale
but the argument still holds because the scale for numerical dissipation
depends on the grid resolution.

The growth rate given by Equation~\eqref{eq:chandra-khi} depends only on $k$
because there are no other length scales in the problem. Introducing a
smoothing length, $a$, makes the growth rate
dependent on this scale. However, smooth profiles have, as previously
mentioned, no simple expression for the growth rates.  For a
three-zone model, in which the velocity has a linear profile between
$z=-a$ and $z=a$ (see the orange solid line the upper panel of
Fig.~\ref{fig:smooth_growth_illu}), an analytic expression can however be
obtained as \citep{drazin2004hydrodynamic} \be \omega^2 =
\f{V^2}{4 a^2} \left[(2ka-1)^2 - e^{-4ka}\right]
    \label{eq:disp-3zone} \ ,
\en
where $2a$ is the scale on which the velocity changes sign. The growth rate
found using Equation~\eqref{eq:disp-3zone} is shown in the lower panel of
Fig.~\ref{fig:smooth_growth_illu} with an orange solid line. In contrast to the
discontinuous profile,\footnote{Formally, the discontinuous profile has $a=0$ but there is no mathematical 
difficulty associated with the blue curve in the lower panel in Fig.~\ref{fig:smooth_growth_illu} as the slope is independent of the value of $a$, see Equation~\ref{eq:chandra-khi}.} the growth rate does not grow without bound. Instead,
wavenumbers with $k a \gtrsim 0.6$ are stable. These high
wavenumbers are not subject to the KHI because they correspond to wavelengths
that are smaller than the scale on which the velocity changes sign.

The velocity profile in the three-zone model is continuous but its
first derivative is discontinuous. For numerical simulations, a
velocity profile of the form $\varv_x(z) = V \tanh(z/a)$ where $a$ is
the smoothing length (shown with a green solid line in the upper panel
of Fig.~\ref{fig:smooth_growth_illu}), has the advantage that both, the
velocity and its derivatives are continuous.
Analytical progress is however difficult\footnote{A lot of analytical progress
has in fact been made for the $\tanh(z)$ profile, see
\citet{Blumen1970}, section 104 in \citet{Chandrasekhar1961},
section 22 in \citet{drazin2004hydrodynamic} and the references therein.
The analytical studies however mainly determine the stability criterion, i.e.,
the growth rate of the KHI is not found as part of the analysis.
}, and instead we must resort to
solving Rayleigh's equation numerically \citep{Michalke1964}. Here
Rayleigh's equation, an eigenvalue equation derived from the fluid
equations in the incompressible limit, is given by
(\citealt{Rayleigh1879}, see e.g., \citealt{Pringle2007}) \be
\left(\f{\omega}{k} - \varv_x\right) \left(\f{d^2 \dvz}{dz^2} - k^2
\dvz\right) + \f{d^2\varv_x}{dz^2} \dvz = 0 \ ,
\label{eq:Rayleigh}
\en
where $\dvz(z)$ is the perturbed velocity in the $z$-direction.
We solve Equation~\eqref{eq:Rayleigh} on $z\in [-\infty, \infty]$ by using \textsc{psecas}
and assume that $\delta \varv_z$ goes to zero at $z\rightarrow \pm \infty$. The
resulting growth rates, originally obtained by \citet{Michalke1964},
are shown with a green solid line in the lower panel of
Fig.~\ref{fig:smooth_growth_illu}. The qualitative features are identical to the
three-zone model, i.e., introducing a smoothing length gives a $k_
{\mathrm{cut}}$ above which the KHI does not grow. Introducing a smoothing
length thus allows numerical simulations of the linear regime to converge.

The above discussion considered an incompressible flow on the infinite
$z$-domain. Numerical astrophysical simulations, and the systems that they
model, are however compressible, and often limited in spatial extent. This
motivates introducing the planar slab shown in Fig.~\ref{fig:khi_setups}. The
planar slab we consider is an over-dense central region with density $\rho_0(1
+\delta)$ moving to the right surrounded by regions with density
$\rho_0$ moving to the left. The planar slab introduces an extra length scale
to the problem, i.e., the width of the slab, $L$.
For discontinuous
density and velocity profiles, the dispersion relation for this setup is
derived in
\citet{Mandelker2016}. The boundary conditions that the
perturbations go to zero as $z\rightarrow\pm \infty$ was assumed and a very
detailed analysis of the dispersion relation was performed. In particular, it
was found that the effective growth rate diverges logarithmically
\citep{Mandelker2016}, i.e.,
$\sigma \propto \ln(k L)$. This means that
the slab with a discontinuous profile is a problematic initial condition for
numerical simulations.

The
planar slab also allows for periodic boundary conditions in the $z$-direction
because
the fluid parameters (the base state) are identical on each side of the slab.
By employing periodic boundary conditions in the $z$-direction, the planar slab
becomes two slabs that are connected to each other on both sides.
It is difficult to imagine a physical system with these properties
and the two connected periodic slabs differ both qualitatively and
quantitatively
from the slab on an infinite domain (see Section~\ref{sec:bodymodes}).
Nevertheless, the double periodic setup is very useful for code testing
purposes if a smooth profile is considered \citep{Lecoanet2016}.
Periodic boundary conditions can be implemented with machine precision and
code testing with periodic boundary conditions thus ensures that
numerical artifacts from the boundaries do not influence the simulation.
Additionally, periodic boundary conditions in the $z$-direction
eliminate the otherwise inevitable errors that would be associated with
truncating the infinite $z$-domain.
We therefore develop the linear theory and perform simulations of the periodic
slab with smooth initial conditions in Section~\ref{sec:simulations}. The linearized
equations that we derive in the next section are however not dependent on this
choice of boundary conditions.

Finally, we will also discuss the KHI as it arises in a cylindrical
stream (see Fig.~\ref{fig:khi_setups}) where the stream has
velocity $V$ ($-V$) inside (outside) a certain radius, $R$.
This configuration can be applied to cold flows in galaxy formation and has
been extensively studied in \citet{Mandelker2016}. \citet{Mandelker2016}
assumed a discontinuous profile for the velocity and density while we
introduce a smoothing length in radius, $a$. We calculate the growth rates
and compare our findings to the results obtained from the analytical
dispersion relation of \citet{Mandelker2016} in Section~\ref{sec:body_cyl}.

\section{Equations and Equilibria}
\label{sec:equations_and_equilibria}

We consider three different sets of equations, i.e., the equations of
inviscid hydrodynamics (e.g., \citealt{batchelor2000introduction}),
the equations of ideal MHD
(e.g., \citealt{Freidberg2014})
and the equations of
Braginskii-MHD \citep{Bra,schekochihin_plasma_2005}.
Braginskii-MHD, an extension of ideal MHD
in which anisotropic transport of heat and momentum is described by the
coefficients $\chi_\para$ and $\nu_\para$, reduces to ideal MHD in the limit
$\chi_\para=\nu_\para=0$. In addition, ideal MHD reduces to inviscid
hydrodynamics by setting $\vec{B}=\bs{0}$ in the equations. We thus introduce the
equations of Braginskii-MHD and consider its various limits. The
mass continuity equation, momentum equations, induction equation and entropy
equation are given in SI units by
\begin{align}
    \pder{\rho}{t} &= -\bs{\nabla} \bs{\cdot} \left(\rho \vec{\varv}\right) \ ,
    \label{eq:rho}
\end{align}
\begin{align}
    \rho \der{\vec{\varv}}{t} &=
    - \bs{\nabla} p -\bs{\nabla} \bs{\cdot} \mathbf{\Pi}
    - \bs{\nabla} \bs{\cdot} \left(\f{B^2}{2\mu_0}\mathbf{1} -
    \f{B^2}{\mu_0}\b\b \right) \ ,
    \label{eq:mom}
\end{align}
\begin{align}
    \pder{\vec{B}}{t} &= \bs{\nabla} \bs{\times} \left(\vec{\varv} \bs{\times} \vec{B}\right)
    \label{eq:ind}\ ,
\end{align}
\begin{align}
    \f{p}{\gamma -1}\der{\ln \left(p \rho^{-\gamma}\right)}{t} &=
    -\mathbf{\Pi} \bs{:} \bs{\nabla} \vec{\varv} - \bs{\nabla} \bs{\cdot} \vec{Q}\ ,
    \label{eq:ent}
\end{align}
where $\bs{a}\bs{b}$ is the dyadic product of vectors $\bs{a}$ and
$\bs{b}$, $\rho$ is the mass density, $\vec{\varv}$ is the mean fluid
velocity, $p$ is thermal pressure, $\vec{B}$ is the magnetic field
with local direction $\b$, $\mu_0$ is the magnetic permeability and
$\gamma=5/3$ is the adiabatic index.

In Equations~\eqref{eq:rho} to \eqref{eq:ent}, the extra terms
that are included in Braginskii-MHD, compared to the equations of ideal MHD,
are the anisotropic heat flux,
$\vec{Q}$, and the anisotropic viscosity tensor, $\mathbf{\Pi}$.
These terms are included in order to model a plasma which is magnetized and
weakly collisional\footnote{See e.g. \citealt{Bal00,Bal01,Qua08,Squire2016}
and references thereto for some of the interesting effects these terms can
cause.}.
The anisotropic heat flux is given by
\be
    \vec{Q} = - \chi_\para \b (\b \bs{\cdot} \bs{\nabla} T) \ ,
\en
where $\chi_\para$ is the heat conductivity and  the anisotropic viscosity
tensor is given by
\be
    \mathbf{\Pi} = - \rho\nu_\para\left(3\b\b\vec{:}\bs{\nabla} \vec{\varv}
     -\bs{\nabla}\bs{\cdot}\vec{\varv}\right) \left(\b\b - \frac{\mathbf{1}}{3}\right) \ ,
    \label{eq:Pi}
\en
where $\nu_\para$ is the viscosity coefficient. The anisotropic transport
coefficients are assumed to be given by their Spitzer values
\citep{Spitzer1962,Bra}, i.e.,
$\chi_\para = \chi_{\para, 0} (T/T_0)^{5/2}$ and $\nu_\para = \nu_{\para, 0}
\,\rho_0/\rho\,(T/T_0)^{5/2}$, where $\chi_{\para,0}$ and $\nu_{\para,0}$
are reference values.

\subsection{Linearized Equations}
\label{sec:linearized-equations}

We consider an equilibrium magnetic field and velocity which is in the
$x$-direction and have magnitudes that can vary along $z$, i.e., given by
$\vec{B} = B(z) \ex$ and $\vec{\varv} = \varv(z) \ex$. We also allow the
density, temperature and
pressure to vary with $z$, i.e. $\rho(z)$, $T(z)$ and $p(z)$ with $p(z)=\kb
\rho(z) T(z)/\mu \mH$ where $\kb$ is Boltzmann's constant, $\mH$ is the proton
mass and $\mu$ is the mean molecular weight. The only
requirement on $B(z)$ and $p(z)$ is that equilibrium demands
\be
    \der{}{z} \left(\f{B^2}{2\mu_0} + p\right) = 0 \ .
    \label{eq:equilibrium-condition}
\en We linearize Equations~\eqref{eq:rho} to \eqref{eq:ent} in order to study the
linear stability properties of a given equilibrium profile. We use a
Fourier transform in the $x$-direction but retain the derivatives in
the $z$-direction. A Fourier transform cannot be used in the
$z$-direction because of the non-trivial $z$-dependence of the
equilibrium.  The perturbations have the form $f_k(z)\exp(ik x -
i\omega t)$ where $f(z)$ is a $z$-dependent Fourier amplitude of a
given variable. We assume that the solutions are constant along the
$y$-direction.  In effect, this corresponds to linearizing the
equations with\footnote{For notational simplicity, we have omitted a
  tilde symbol on top of Fourier amplitudes. It should still be clear
  from the context when we are using Fourier amplitudes.}
\begin{align}
    \rho &\rightarrow \rho(z) + \delta \rho(x, z) \ , \\
    T &\rightarrow T(z) + \delta T(x, z) \ , \\
    \vec{\varv} &\rightarrow \varv(z) \ex + \delta \vec{\varv}(x, z) \ , \\
    \vec{B} &\rightarrow B(z) \ex + \delta \vec{B}(x, z) \ ,
\end{align}
and performing the Fourier transform in time and in the $x$-direction.

We introduce a vector potential $\vec{A} = A(z) \ey$ such that the
magnetic field is given by $\vec{B} = \bs{\nabla} \bs{\times} \vec{A}$. It
follows that the perturbation to the magnetic field is given by
$\delta \vec{B} = \bs{\nabla} \bs{\times} (\delta A \ey)$
such that
\begin{align}
    \delta B_x &= -\pder{\delta A}{z} \ , \\
    \delta B_z &= i k \delta A \ .
\end{align}
The advantage gained by introducing the vector potential is that the induction
equation is reduced to a single equation (instead of two).
We obtain a set of linearized equations given by the continuity
equation,
\begin{align}
    %%%%%%%%%%%%%%%%%%%%%%%%%%%%%%%%%%%%%%%%%%%%%%%%%%%%%%%%%% Continuity
    - i \omega \drho &=  - i k \left(\varv \drho + \dvx \right)
    - \left(\der{\ln \rho}{z} + \pder{}{z} \right) \dvz \ ,
    \label{eq:rho-lin}
\end{align}
the $x$-component of the momentum equation,
\begin{align}
    %%%%%%%%%%%%%%%%%%%%%%%%%%%%%%%%%%%%%%%%%%%%%%%%%%%%%%%%%%% x-momentum
    -i \omega \dvx &=
    % advection
    -ik \varv \dvx - \pder{\varv}{z} \dvz
    % pressure
    - i k \cs^2 \left(\drho + \dt\right)
    % lorent force
    + ik\va^2\der{\ln B}{z} \dA
    % viscosity
    \nonumber \\
    &-\nu_\para \left(\f{4}{3}  k^2 \dvx + 2  k^2 \pder{\varv}{z} \dA +
    \f{2}{3}  i k \pder{\dvz}{z} \right) \ ,
    \label{eq:mom-x-lin}
\end{align}
the $z$-component of the momentum equation,
\begin{align}
    %%%%%%%%%%%%%%%%%%%%%%%%%%%%%%%%%%%%%%%%%%%%%%%%%%%%%%%%%%% z-momentum
    -i \omega \dvz &=
    % advective
     -i k \varv \dvz
    % pressure
    - \cs^2 \pder{}{z} \left(\drho + \dt\right)
    + \cs^2 \der{\ln p}{z} \left(\drho - \dt\right)
    % lorentz
    \nonumber \\ &
    +\f{\va^2}{B}\left[
    \pdder{}{z} +
    \der{\ln B}{z}\pder{}{z}
    - k^2
    \right] \delta A
    % viscosity
    \nonumber \\ &
    \: -\nu_\para\left(\f{5}{2} \der{\ln T}{z} + \pder{}{z} \right)
     \left(
    \f{2}{3} i k \dvx + ik \pder{\varv}{z} \dA - \f{1}{3}\pder{\dvz}{z}
    \right) \ ,
    \label{eq:mom-z-lin}
\end{align}
an equation for the magnetic vector potential,
\begin{align}
    %%%%%%%%%%%%%%%%%%%%%%%%%%%%%%%%%%%%%%%%%%%%%%%%%%%%%%%%%%% induction
    -i \omega \dA &= \dvz - ik \varv \dA \ ,
    \label{eq:ind-lin}
\end{align}
and the entropy equation,
\begin{align}
    %%%%%%%%%%%%%%%%%%%%%%%%%%%%%%%%%%%%%%%%%%%%%%%%%%%%%%%%%%% entropy
    -i\omega\dt
    &= - ik \left(\varv \dt  + \f{2}{3} \dvx \right)
    - \left(\der{\ln T}{z} + \f{2}{3} \pder{}{z} \right) \dvz
    \nonumber \\ &
    -\f{2k^2\chi_\para T}{3p} \left(\dt + \der{\ln T}{z} \dA \right) \ .
    \label{eq:ent-lin}
\end{align}
Here we have introduced the squared Alfv{\'e}n speed
\be
    \va^2 = \f{B^2}{\mu_0 \rho} \ ,
\en
and the squared isothermal sound speed
\be
    \cs^2 = \f{p}{\rho} \ ,
\en
which define the plasma-$\beta$
\be
    \beta \equiv \f{2 \cs^2}{\va^2} \ .
\en
Equations~\eqref{eq:rho-lin} to \eqref{eq:ent-lin} constitute an eigenvalue problem
that we solve using the Python framework \textsc{psecas}
which we describe in the next section.

\subsection{\textsc{psecas}: A Python package for pseudo-spectral eigenvalue
problems in (astrophysical) fluid dynamics}
\label{sec:psecas}

\begin{figure*}
\includegraphics[trim= 0 25 0 0]{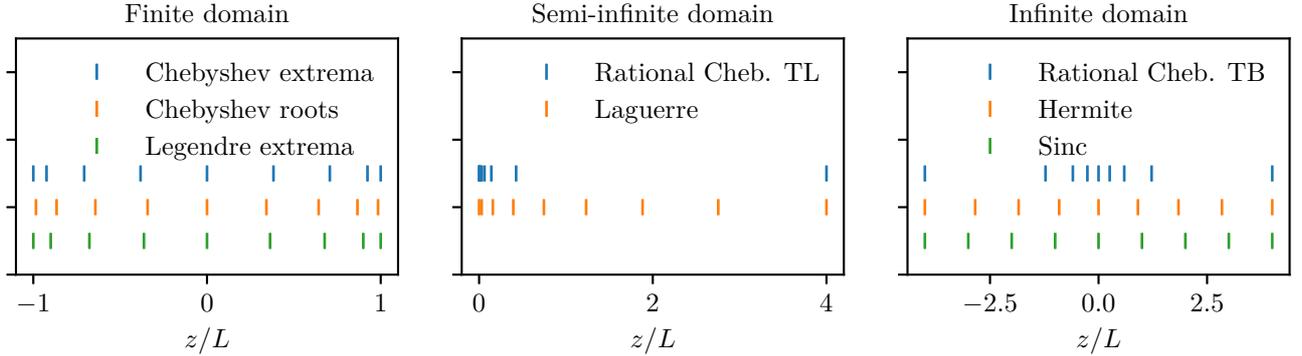}
\caption{Illustration of eight of the nine grids implemented in
\textsc{psecas} with $N=9$ grid points.
The grids used for finite domains have the grid points closely packed
at the boundaries (left panel) while the semi-infinite (middle panel)
and infinite grids (right panel) have their grid points
packed close to the origin. For the grids used for finite domains, the
packing of grid points prevents high frequency
oscillations close to the boundaries. For the (semi)-infinite grids, the
placement of grid points helps with obtaining good spatial resolution at the
region of interest while covering a large extent that mimics the (semi)-infinite
domain. One exception is the Sinc grid which is equally spaced and which
therefore requires many grid points to model the infinite domain. The last
grid available in \textsc{psecas} is the periodic Fourier grid which has a
uniform spacing (not shown). More details can be found
in Table~\ref{tab:grid_table} and in \citet{Boyd}.}
\label{fig:grids}
\end{figure*}

\textsc{psecas} is a newly developed Python package for using pseudo-spectral
methods to semi-automatically generate and solve eigenvalue problems.
The motivation for writing this package is to facilitate solving
the complicated eigenvalue problems that often arise in (astrophysical) fluid
dynamics. Examples of such eigenvalue problems (besides the ones studied in
this paper) include the quasi-global theory of the heat-flux-driven buoyancy
instability \citep{Lat12}, the heat-and-particle-flux driven buoyancy
instability  \citep{Berlok2016b}, the magnetorotational instability with the
Hall effect (Hall-MRI, \citealt{Bethune2016,Krapp2018}),
the thermal instability in
galaxy clusters \citep{Choudhury2016} and the vertical shear instability
\citep{Umurhan2016}.

\textsc{psecas} works by generating a matrix eigenvalue problem from a
linearized set
of equations by using the pseudo-spectral method
to discretize the equations (see e.g.,
\citealt{fornberg1996practical,trefethen2000spectral,Boyd}).
Users can enter their equations as strings which
are then parsed by the program, allowing a close resemblance
between code and the equations as they appear on paper.\footnote{This
was inspired by the popular code \textsc{Dedalus}
\citep{Dedalus} which shares several features with \textsc{psecas}.} This
reduces the risk
of introducing errors when solving complicated linearized equations such as
Equations~\eqref{eq:rho-lin} to \eqref{eq:ent-lin}.

The key ingredient in pseudo-spectral methods is the grid and the
corresponding polynomials used to discretize the problem. \textsc{psecas}
contains several options for solving problems on a periodic interval (e.g.,
$z \in [0,\, 2\pi]$),
a finite interval $z \in [0,\, L]$, a semi-infinite interval $z \in [0,\,
\infty]$ or an infinite interval $z \in [-\infty,\, \infty]$. These are
summarized in Fig.~\ref{fig:grids} and in Table~\ref{tab:grid_table}.
In astrophysical fluid dynamics, the grids on the semi-infinite domain will
typically be used to represent the radial coordinate in spherical or
cylindrical polar coordinates. The grids on the infinite domain are useful
for e.g., representing the distance above the midplane in a galactic disk or
accretion disk. In this paper we use the Fourier grid for the
planar slab on a periodic domain, the rational Chebyshev TB grid for the
planar slab on an infinite domain and the rational Chebyshev TL grid for the
cylindrical stream. The finite grids are useful for studying
e.g., plane-parallel atmospheres and can also be used for the KHI with
reflective boundaries.\footnote{We have used the Chebyshev extrema grid to
reproduce Figure 4 in \citet{Miura1982}. A script for performing this
calculation is included in \textsc{psecas}.}

Given a set of $d$ linearized equations and a specified grid with $N$ grids
points \textsc{psecas} automatically generates a (generalized) matrix
eigenvalue problem of dimension $N d\times N d$. The matrix eigenvalue
problem is solved using \textsc{Scipy} \citep{scipy} which returns $N$
eigenvalues most of which will not be physical but numerical \citep{Boyd}.
An important aspect of solving eigenvalue problems with pseudo-spectral
methods is therefore ensuring that the computed eigenvalues have converged and
are physical (\citealt{Boyd}, definition 16). \textsc{psecas} has
functionality for iteratively increasing the number of grid points until the
obtained growth rates differ by less than a user-specified tolerance with
respect to the previous iteration. For large $N$, which might be required for
convergence, solving the full eigenvalue problem can be computationally
intensive.
For instability studies, one is however often only interested in the
fastest growing eigenvalues.
To take advantage of this, \textsc{psecas} contains functionality for only
finding a single eigenvalue in the vicinity of a guess. This functionality is
used for iteratively finding converged solutions, i.e., by using the result
from a previous iteration as a guess for the next iteration at higher grid
resolution. To further enable large parameter studies, \textsc{psecas} uses
the message passing interface (MPI) to distribute calculations to several
hundred processors \citep{Dalcin2008}.

A common task for instability studies is finding the maximal growth rate of
the instability. \textsc{psecas} uses golden section search
(\citealt{press2007numerical}, section 10.2) for finding the maximum of an
instability with respect to a given parameter. This facilitates finding the
maximally growing wavenumber to a specified precision, e.g., $10^{-8}$ for the
calculations presented in Table~\ref{tab:khi_table}.

\subsection{Equilibria}
\label{sec:equilibria}

We consider the periodic slabs (see Fig.~\ref{fig:khi_setups})
with an equilibrium velocity profile given by \citep{Lecoanet2016}
\begin{align}
    \f{\varv(z)}{V} =  \tanh\left(\f{z-z_1}{a}\right) -
                       \tanh\left(\f{z-z_2}{a}\right) - 1 \ ,
    \label{eq:v-equi}
\end{align}
where $z_1=L/2$, $z_2=3L/2$ and $a = 0.05L$ is a smoothing length.
This profile is periodic on the domain $z=[0, 2L]$ and changes direction at
$z_1$ and $z_2$ where the KHI can be triggered. The magnitude of the velocity
is set by the free parameter $V$. Following \citet{Lecoanet2016}, we
also allow for a density variation
\begin{align}
    \f{\rho(z)}{\rho_0} =
    1 + \f{\delta}{2} \left[\tanh\left(\f{z-z_1}{a}\right) -
                       \tanh\left(\f{z-z_2}{a}\right)\right] \ ,
    \label{eq:rho-equi}
\end{align}
with a magnitude set by the parameter $\delta$. With $\delta=0$ the
background density is uniform. Equations~\eqref{eq:rho-lin} to \eqref{eq:ent-lin}
allow for general variations of pressure, $p(z)$, and magnitude of the
magnetic field strength, $B(z)$, as long as
Equation~\eqref{eq:equilibrium-condition} is fulfilled. Here we consider the
simplest configuration of this type, where both background pressure
and magnetic field strength are constant. This yields the temperature
variation \be \f{T(z)}{T_0} = \f{\rho_0}{\rho(z)} \ , \en and an
initial magnetic field which is completely specified by the
plasma-$\beta$. Anisotropic transport does not affect the
equilibrium,\footnote{This is in contrast to isotropic diffusion.  The
  inclusion of isotropic viscosity would change the velocity profile
  and isotropic heat conduction changes the temperature profile when
  $\delta\neq0$.}  which has zero heat and momentum flux regardless of
the values of $\chi_\para$ and $\nu_\para$ when the magnetic field is
in the $x$-direction.

\begin{table}
    \centering
    \caption{The nine grids available in \textsc{psecas}.}
    \label{tab:grid_table}
    \begin{tabular}{llll}
        \hline
        Domain & Grid & Reference\\
        \hline
        Periodic \\
        & Fourier  & \citet{trefethen2000spectral}\\
        Finite \\
        & Chebyshev extrema  & \citet{Boyd} F.8\\
        & Chebyshev roots  & \citet{Boyd} F.9\\
        & Legendre extrema  & \citet{Boyd} F.10\\
        Semi-infinite \\
        & Rational Chebyshev, TL  & \citet{Boyd1987a}\\
        & Laguerre  & DMSuite \\
        Infinite \\
        & Rational Chebyshev, TB  & \citet{Boyd1987}\\
        & Hermite grid  & DMSuite \\
        & Sinc grid  & \citet{Boyd} F.7 \\
        \hline
    \end{tabular}
\end{table}

\begin{table}
    \centering
    \caption{Parameters for the four examples of the KHI.
    Listed are also the fastest growing wavenumbers and
    the corresponding growth rates.}
    \label{tab:khi_table}
    \begin{tabular}{lllllll}
        \hline
        Name & $\beta^{-1}$ & $\delta$ & $\nu_\para/(L c_0)$
        & $k_{\mathrm{max}}L$ & $\sigma_{\mathrm{max}}L/c_0$ \\
        \hline
        H & 0 & 0 & 0 & 5.1540899 & 1.7827486 \\
        H$\delta$ & 0 & 1  & 0 & 3.5128319 & 1.4035133\\
        M & $0.2$ & 0 & 0 &  5.5775520 & 1.4614214 \\
        M$\nu$ & $10^{-3}$ & 0 & 0.01  & 4.5470431 & 1.7087545 \\
        \hline
    \end{tabular}
\end{table}

\begin{figure}
\includegraphics[trim= 0 20 0 0]{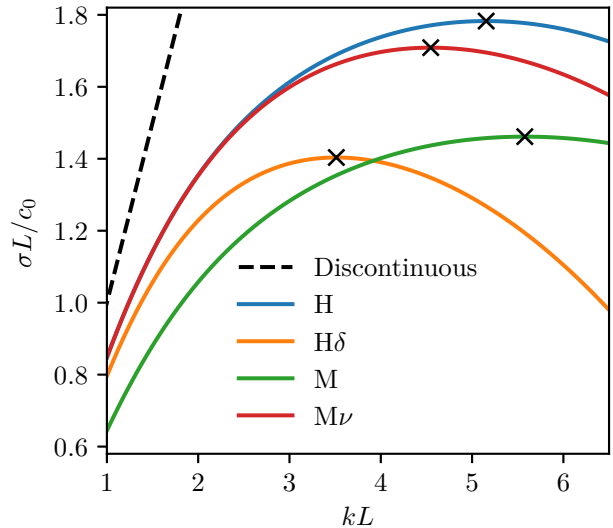}
\caption{Growth rates as a function of wavenumber for the doubly periodic
domain and the parameters listed in
Table~\ref{tab:khi_table}. Magnetic tension, viscosity and a density variation all
inhibit the KHI and a density variation also turns the KHI into an
overstability. A smooth transition on scale $a=0.05L$ inhibits growth at all scales visible in the plot
and stabilizes the KHI at high wavenumbers. This is in contrast to the 
incompressible, analytic solution for a piece-wise constant velocity profile,
which grows without bound (dashed black line, Equation~\ref{eq:chandra-khi} with $\delta=0$).}
\label{fig:growthrates}
\end{figure}

Using \textsc{psecas}, we discretize Equations~\eqref{eq:rho-lin} to \eqref{eq:ent-lin} on a
periodic Fourier grid in $z$ and approximate the differential operators using
differentiation matrices. In this basis, periodic boundary conditions are
automatically applied to the perturbations and the resulting matrix eigenvalue
problem has dimension $5N\times5N$ where $N$ is the number of grid points
which is iteratively increased.

We consider four different cases
which are chosen
in order to test different parts of \textsc{Athena}.
The four cases are hydrodynamics with a uniform
density (H), hydrodynamics with a varying background density (H$\delta$),
MHD with a strong magnetic field (M) and Braginskii-MHD with a weak magnetic
field and anisotropic viscosity (M$\nu$). We set the flow velocity to be
$V/c_0 = 1$ in all four cases and summarize their other parameters
in Table~\ref{tab:khi_table}. Here $c_0 = \sqrt{p_0/\rho_0}$ is the 
isothermal sound speed at temperature $T_0$, which is related to the 
isothermal sound speed in the  denser slab by $c_s = c_0/\sqrt{\delta + 1}$.

The resulting growth rates are shown in
Fig.~\ref{fig:growthrates} where we have defined $\sigma \equiv -\mathrm{Im}
(\omega)$. All of the calculations with $\delta = 0$ have
$\mathrm{Re}(\omega)=0$, i.e., they are purely growing instabilities.
Only the KHI with a density variation, $\delta\neq0$, has
$\mathrm{Re}(\omega)\neq0$, and we find that $
\mathrm{Re}(\omega)=\pm0.5422067$ when $k=k_{\mathrm{max}}$. This corresponds
to an
oscillating or a
traveling growing mode\footnote{Depending on the excitation in the same way
that e.g., sound waves can be traveling or standing.}.
The fastest growing mode for each
set of physical parameters is indicated with black crosses in
Fig.~\ref{fig:growthrates}. The corresponding values of $k_{\mathrm{max}}$
and $\sigma_{\mathrm{max}}$ are given in Table~\ref{tab:khi_table}. These values
have been calculated within a tolerance of $10^{-8}$.
It is clearly seen in Fig.~\ref{fig:growthrates} and Table~\ref{tab:khi_table} that
the fastest growing
instability is found when the background density is uniform (blue) and that
the instability is inhibited when a density
variation is introduced. This qualitatively agrees with the expectation from
Equation~\eqref{eq:chandra-khi}, derived in
the incompressible limit for the planar sheet \citep{Chandrasekhar1961}.
The KHI with a uniform background is
inhibited by Braginskii viscosity \citep{Suzuki2013} and by magnetic tension
\citep{Chandrasekhar1961,Miura1982}.
We observe that Braginskii viscosity does not damp the long wavelengths
as viscosity is ineffective on those scales for the chosen value of the
viscosity parameter (magnetic tension is negligible for this
calculation).

Besides the eigenvalues, \textsc{psecas} also returns the eigenvectors
of the eigenvalue problem, i.e., numerical solutions for $\delta \rho/\rho$,
$\delta T/T$, $\delta \varv_x$, $\delta \varv_z$ and $\delta A$. These can be
used to construct the two-dimensional linear solutions of the initial value
problem. We use the obtained two-dimensional linear solutions to
initialize simulations with \textsc{Athena} in the following section.

\section{Simulations of surface modes}
\label{sec:simulations}

\begin{figure*}
\includegraphics{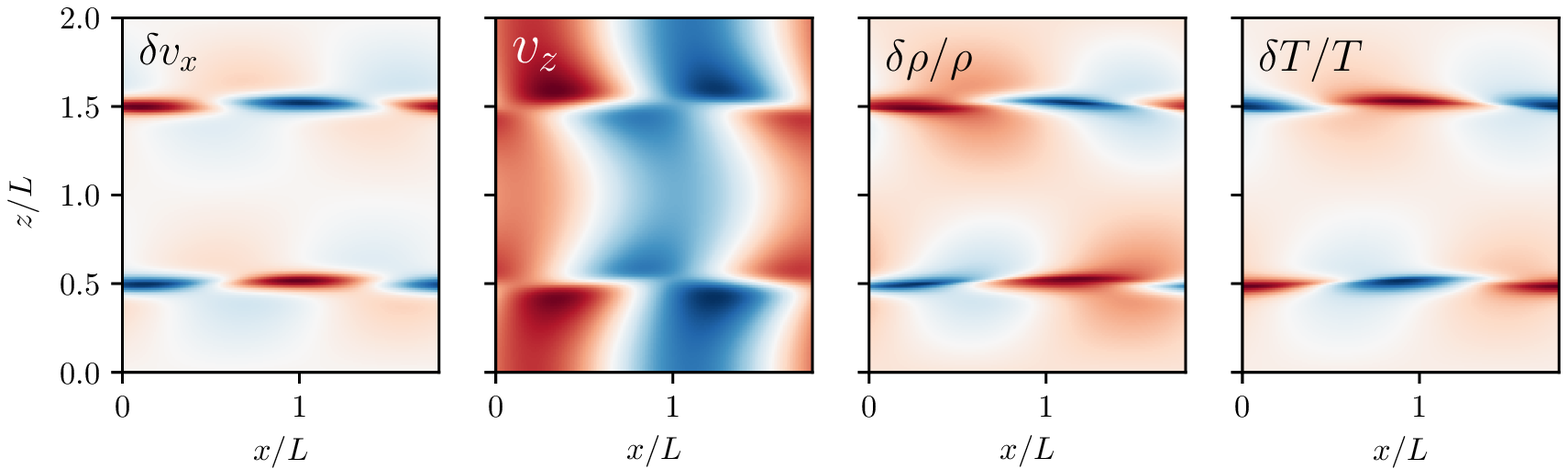}
\includegraphics[trim=0 15 0 0]{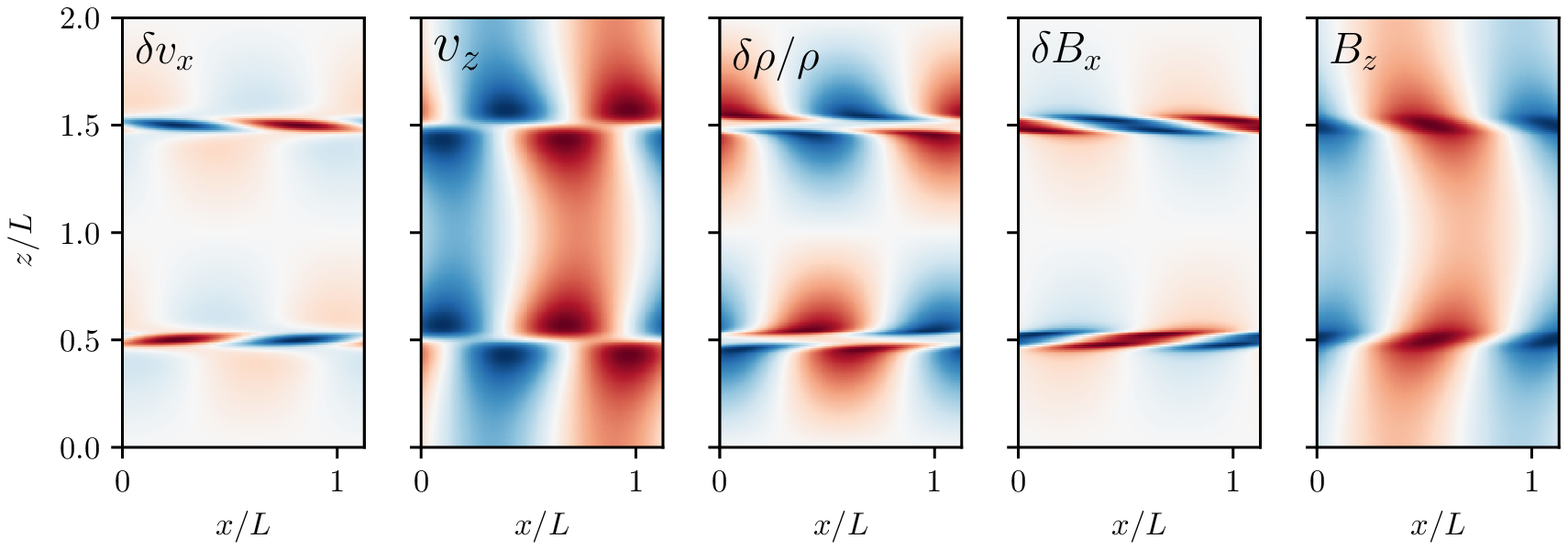}
\caption{Deviations from the background functions, Equations~\eqref{eq:v-equi}
  and \eqref{eq:rho-equi}, in two of the \textsc{Athena} simulations. The
  shear-interfaces are located at $z/L=1/2$ and $z/L=3/2$ giving rise to
  pronounced disturbances at those locations. All images are visually
  indistinguishable from the linear solution given by
  Equation~\eqref{eq:lin_solution}.  \emph{Upper row:} Simulation of the KHI
  with a background density variation ($\delta=1$). The modes are exponentially
  growing while moving to the right at a constant speed.  \emph{Lower row:}
  Simulation of the KHI with a magnetic field $\vec{B}=B\ex$ which inhibits the
  instability ($\beta=5$). The modes are purely growing.}
\label{fig:eig_2d}
\end{figure*}

We use the publicly available MHD code \textsc{Athena} to perform the tests
\citep{gardiner_unsplit_2005,Gardiner2008,stone_athena:_2008}.
We consider a two-dimensional
domain of size $L_x\times L_z$ where $L_z=2L$ and $L_x = 2\pi/k_{\mathrm{max}}$
where $k_{\mathrm{max}}$ is the wavenumber for the fastest growing mode of
the KHI for the given parameters
(see Table~\ref{tab:khi_table}). We set $L=\rho_0=p_0=1$ as code units
and excite the instability by using the linear solutions found in the previous
section.
When initializing the simulations, each component of the perturbation,
$\delta f(x, z, t)$ in real space, is related to its $z$-dependent Fourier
component, $f_k(z)$, by
\begin{align}
    \delta f(x, z, t) = 2A\,\mathrm{Re}\left[f_k(z) e^{-i\omega t + i k
    x}\right]
    \ ,
    \label{eq:lin_solution}
\end{align}
where $A=10^{-4}$ is the amplitude\footnote{The eigenmodes are normalized such that the maximum
magnitude of the real and imaginary parts of the velocity components is 1.} of the
perturbation. The eigenmodes returned by \textsc{psecas} are given at 
the grid points illustrated in Figure~\ref{fig:grids} but can equivalently be 
expressed in terms of the corresponding basis functions, i.e.,
\begin{align}
    f_k(z) = \sum_l c_l P_l(z) \ ,
    \label{eq:basis-representation}
\end{align}
where $P_l(z)$ is a set of basis functions and $c_l$ are the corresponding
coefficients \citep{Boyd}.
For the Fourier grid considered here, Equation~\eqref{eq:basis-representation}
is simply a Fourier series and $P_l(z)$ are complex exponentials. For the
polynomial grids, $P_l(z)$ are the corresponding polynomials (e.g., Legendre
polynomials for the Legendre grid). This means that \textsc{psecas}
solutions can be evaluated at any location with spectral accuracy by using 
Equation~\eqref{eq:basis-representation} \citep{trefethen2000spectral}.

We evaluate all perturbed gas quantities (density, pressure and 
velocities) at the grid cell centers 
in \textsc{Athena}.\footnote{\textsc{Athena} is a finite volume code 
and there is therefore an error 
associated with simply evaluating the spectral solution at cell
centers. A more correct treatment would instead initialize the volume
average of the spectral solution over each cell, i.e.,
\begin{align}
    \delta \bar{f}(x, z, t) = (\Delta x \Delta z)^{-1} \int_{z-\Delta z/2}^
    {z+\Delta z/2} 
    \int_{x-\Delta x/2}^{x+\Delta x/2} \delta f(x', z', t) \, dx' \, dz'
\end{align}
The required integral can be performed analytically, e.g., for the Fourier
grid
\be
\delta \bar{f}(x, z, t) = 
2A\,\mathrm{Re}\left[
\sum_l c_l \f{\sin(k_l \Delta
z/2)}
{k_l \Delta z/2}
\f{\sin(k \Delta x/2)}{k \Delta x/2} e^{i k_l z + i k x - i \omega t} \right]
\ ,
\label{eq:finite-volume}
\en
where $c_l$ are complex Fourier coefficients and $k_l = 2 \pi l/ L_z$.
Equation~\eqref{eq:finite-volume} only agrees with Equation~
\eqref{eq:lin_solution} in the limit $\Delta x, \Delta z \rightarrow 0$ and a
series expansion of 
the $\sin$ functions shows that the error introduced by using 
Equation~\eqref{eq:lin_solution} instead of Equation~\eqref{eq:finite-volume} is
$O((\Delta x)^2) + O((\Delta z)^2)$, i.e., 
second order in the grid spacing. As \textsc{Athena} is second order accurate,
the more precise treatment has not been necessary in the present work.
Initialization using Equation~\eqref{eq:finite-volume} would however be required
for simulations with finite volume codes that are higher than second order
accurate.} 
For the MHD 
simulations the magnetic field requires special treatment.
Here we evaluate the vector potential at cell faces instead 
of cell centers. The magnetic field is then initialized using finite differences. 
This ensures that the magnetic field has zero divergence to machine precision.

We perform the simulations on a grid with $N_z=256$ and $N_x = N_z L_x/L_z$
(rounding up to nearest integer). We use 3rd order reconstruction and Corner
Transport Upwind (CTU, \citealt{Colella1990})
with the Harten-Lax-van Leer with contact
Riemann solver for the hydrodynamic
simulations (HLLC, \citealt{toro2013riemann}) and the Harten-Lax-van Leer with contact
and Alfv\'{e}n mode Riemann solver for the
MHD simulations (HLLD, \citealt{Miyoshi2005}).
We set the Courant number to 0.8 and evolve the simulations
until $c_0 t/L= 10$.

In Fig.~\ref{fig:eig_2d} we show snapshots from two of the simulations during the
linear phase of the instability. In the upper row of Fig.~\ref{fig:eig_2d}
we show the difference of all four components of the solution with respect to
the equilibrium at $c_0 t/L =5$ for the simulation with a density variation (H$\delta$). At this
point in time the perturbations have grown by a factor $\sim 10^3$ and the
visual
appearance is indistinguishable from the analytic solution
evaluated using Equation~\eqref{eq:lin_solution}. The mode is moving to the right at
speed $\varv_{\mathrm{ph}} = \mathrm{Re}{(\omega)}/k$.

In the lower panel of Fig.~\ref{fig:eig_2d}, we show similar results for the MHD
simulation with $\beta=5$ (here at $c_0 t/L = 3$ due to the larger growth
rate). For this simulation we also present the $z$-component of the magnetic
field and the perturbation to the $x$-component. This instability is purely
growing and the spatial form of the perturbation is therefore constant in time
during the linear phase of the instability, i.e., it is only the magnitude of
the perturbation that changes.

A notable feature of the modes in Fig.~\ref{fig:eig_2d} are the strong variations
close to the shear layers located at $z_1$ and $z_2$. Away from the surfaces
of the
shear layers, the perturbations are much smaller. For that reason, these modes
are called \emph{surface modes}. At the flow velocity considered, i.e.,
$V/c_0=1$, the solutions found in
Table~\ref{tab:khi_table} and Fig.~\ref{fig:growthrates} are all surface modes.
This is in contrast to \emph{body mode} disturbances which extent throughout
the entire domain. We study the body modes that arise when the flow velocity
is higher, $V/c_0 = 2.5$, in Section~\ref{sec:bodymodes}.

\begin{figure*}
\includegraphics[trim=0 15 0 0]{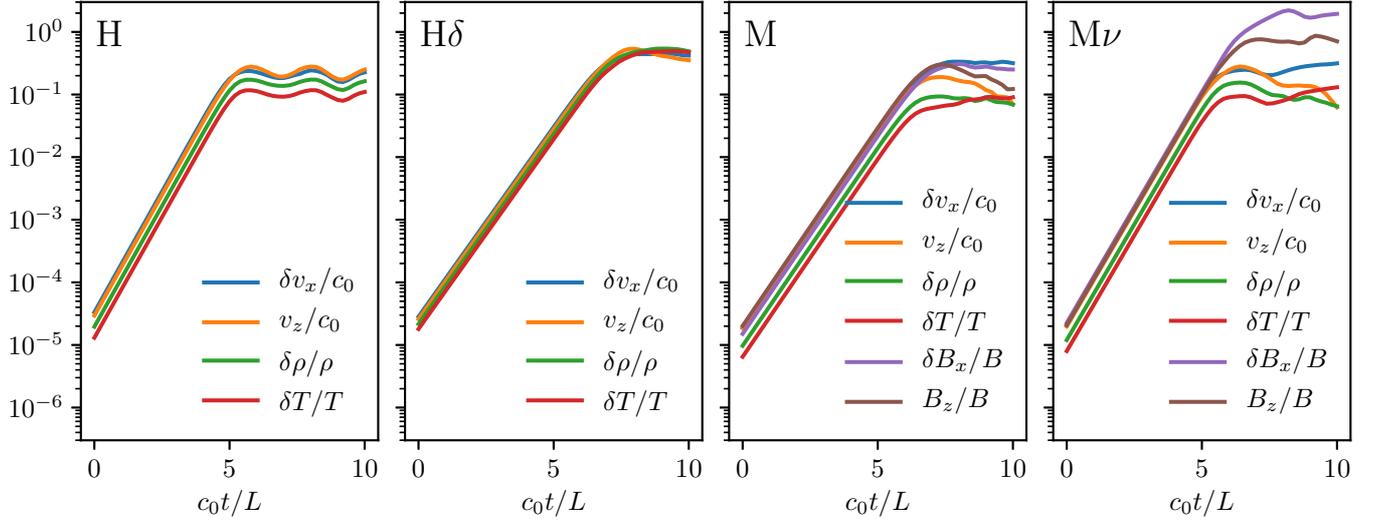}
\caption{Exponential growth of deviations from the background equilibrium
in the four simulations listed in Table~\ref{tab:khi_table}. The exponential growth
starts without an initial transient because the linear solution is used
to excite the instability. The fitted growth rates agree with the theory to
within $5\cdot10^{-3}$ regardless of which component we use
for the fit. Nonlinear effects, not captured by the linear theory, eventually
cause the instability to saturate.}
\label{fig:simulation_growth}
\end{figure*}

\begin{figure*}
\includegraphics[trim = 0 15 0 0]{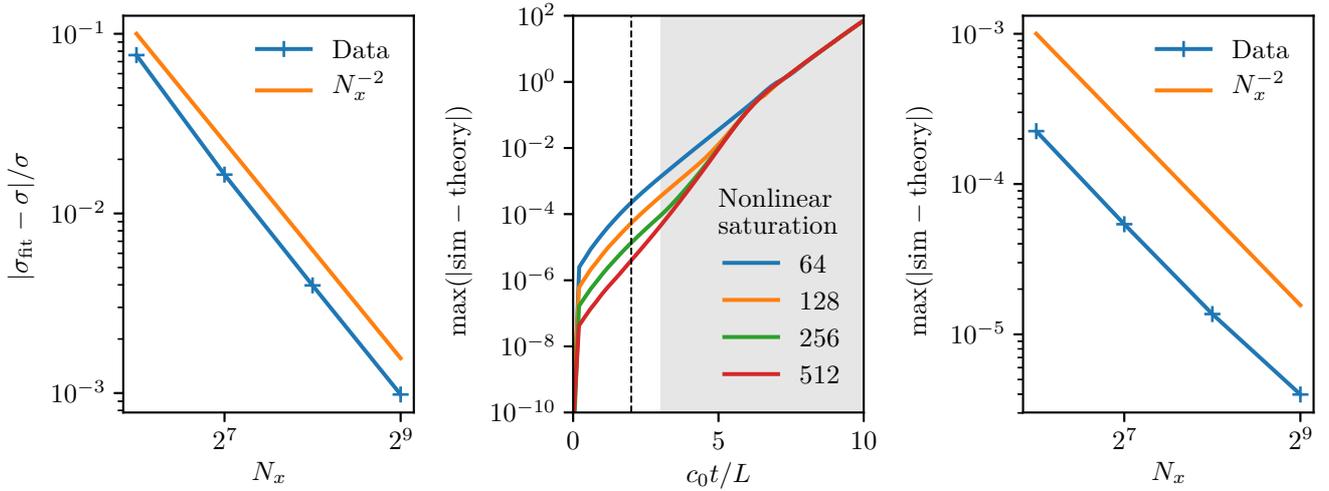}
\caption{Convergence study of \textsc{Athena} simulations to \textsc{psecas}
  solutions for the H$\delta$ model. \emph{Left:} Second order convergence of
  the fractional error in the fit of the growth rate.  \emph{Middle:} Maximum
  difference between linear solution for $\varv_z$ and simulations with
  increasing grid resolution, $N_x$, as a function of time. The gray area
  roughly indicates the region in which nonlinear effects are important. Data
  points from the linear regime (indicated with a black dashed line) are
  extracted and shown in the right panel. \emph{Right:} Second order convergence
  of the maximum difference between the linear solution and simulations at 
  $c_0 t/L=2$.}
\label{fig:error_convergence}
\end{figure*}

For code testing, inspection of the components of the perturbation shown in
Fig.~\ref{fig:eig_2d} can aid in diagnosing errors in the source code by
visually identifying the component in which the error initially occurs. This
type of visual inspection is much more difficult if the instability is not
excited by an eigenmode.  A visual inspection is of course not sufficient for
rigorous code testing.  In Fig.~\ref{fig:simulation_growth} we show the
evolution of perturbations from the background equilibrium for all four
simulations. As expected, we observe a nearly perfect exponential evolution of
all quantities in all four cases.  There is no initial transient and we can
therefore estimate the growth rates for the simulations without ignoring the
first part of the data. This is another advantage compared to simulations where
the instability is seeded with random perturbations or a non-eigenmode analytic
expression. We fit an exponential time evolution of the form $h(t) =
a\exp(\sigma_{\mathrm{fit}} t)$ to the time interval $c_0 t/L =0-2$ and
calculate the relative error of the fitted growth rate as
$|\sigma_{\mathrm{fit}}-\sigma|/\sigma$.  We find agreement with the theoretical
values listed in Table~\ref{tab:khi_table} with a relative error of between
$3\cdot 10^{-3}$ and $5\cdot 10^{-3}$ depending on which of the quantities we
fit. This error should decrease with numerical resolution and can be used for
convergence testing. As an example, we consider hydrodynamic simulations of the
H$\delta$ model with $\delta=1$ at four different resolutions, $N_x=64$, 128,
256 and 512 and show in the left panel of Fig.~\ref{fig:error_convergence} that
the relative error in the growth rate converges at second order.

We can also use the linear solution for convergence testing by directly
measuring the difference between the \textsc{Athena} solution and the linear
solution. We again consider simulations of the H$\delta$ model with $\delta=1$
at four different resolutions, $N_x=64$, 128, 256 and 512. For each simulation
snapshot, we then calculate the maximum value of the absolute difference
between the linear solution for $\varv_z(x, z, t)$ given by
Equation~\eqref{eq:lin_solution} and the simulation snapshot. We show how the
absolute maximum of this difference grows as a function of time in the middle
panel of Fig.~\ref{fig:error_convergence}.

As \textsc{Athena} solves the full nonlinear equations of hydrodynamics,
Equations~\eqref{eq:rho} to \eqref{eq:ent}, the difference grows without bound at late
times. This is because the KHI in the \textsc{Athena} simulation saturates in
the non-linear regime while the linear solution predicts continued exponential
growth.  The difference at large times is thus not due to errors in
\textsc{Athena} but rather due to the neglect of higher order terms in
Equations~\eqref{eq:rho-lin} to \eqref{eq:ent-lin}. At early times ($c_0 t/L \lesssim 3$),
however, where nonlinear terms are small, we do observe a decrease in the
difference between \textsc{Athena} and the linear solution as the numerical
resolution is increased. This difference is due to finite resolution in
\textsc{Athena} and it converges at second order, see the right panel of
Fig.~\ref{fig:error_convergence}.

\section{High-Mach-number flows with body modes}
\label{sec:bodymodes}

\begin{figure*}
\includegraphics[trim = 0 10 0 0]{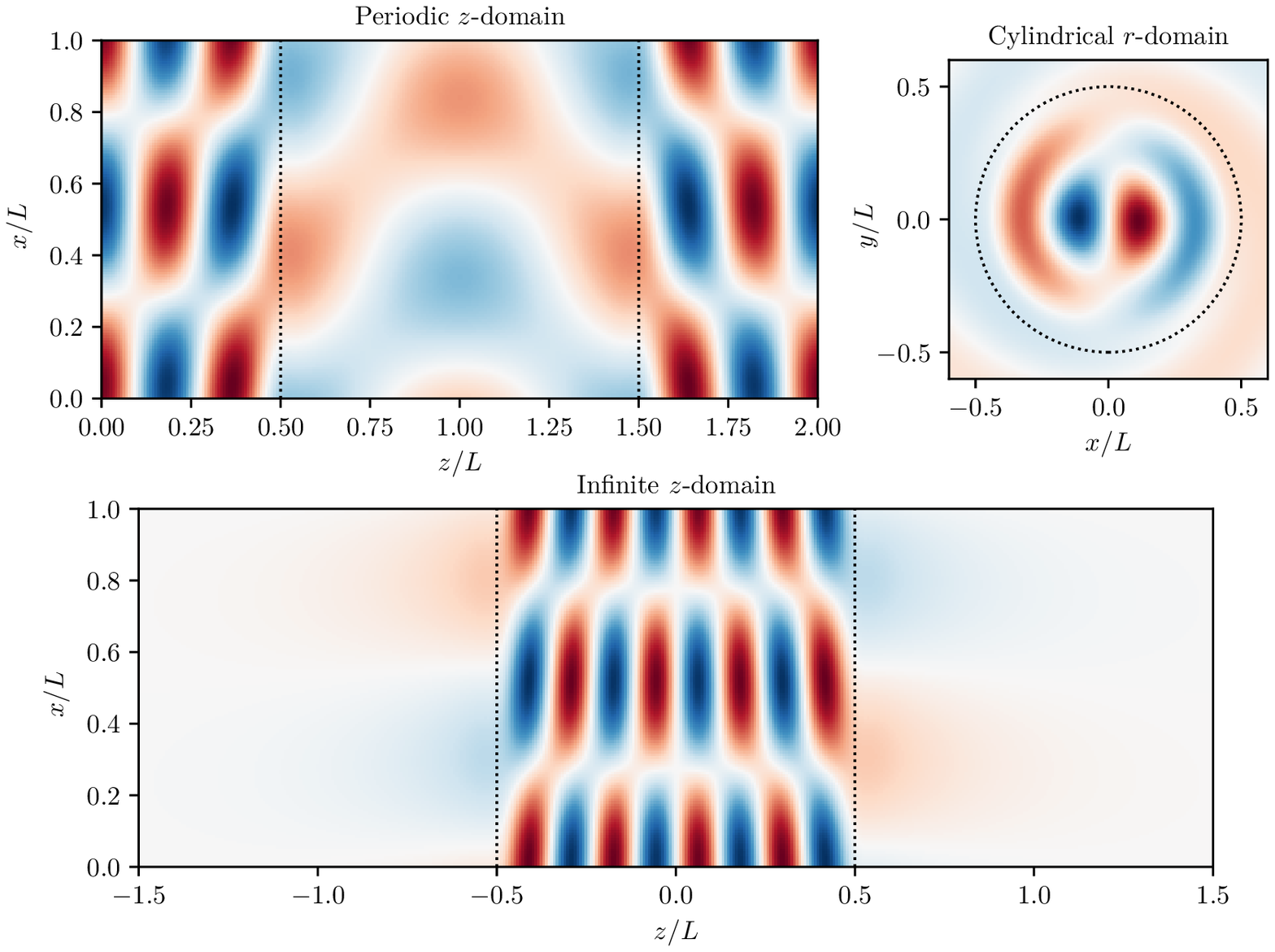}
\caption{ The KHI in three different setups. At high flow speeds
  $V/c_0=2.5$ an over-dense slab ($\rho=2\rho_0$) gives rise to pressure
  profiles that are not confined to the surfaces of the shear interfaces
  (indicated with black dotted lines) but rather extent throughout the interior
  of the slab. These are body modes (or reflective modes) which consist of
  soundwaves that are resonant inside the body of the slab. The periodic
  $z$-domain consists of two connected slabs with body modes inside both of
  them, and the pressure perturbation thus extends the entire domain.  This is
  in contrast to the infinite $z$-domain and the cylindrical $r$-domain where
  the perturbations decay with distance from the surfaces of the slab. The
  calculations have been performed with \textsc{psecas} and a smoothing length
  $a=0.05L$. The Cartesian calculations show modes with $k=k_x=2\pi/L$ while the
  cylindrical calculation used $k=k_z=2\pi/L$ and azimuthal wavenumber $m=1$
  (dipole).}
\label{fig:body_modes}
\end{figure*}
\begin{figure*}
\includegraphics[trim = 0 25 0 0]{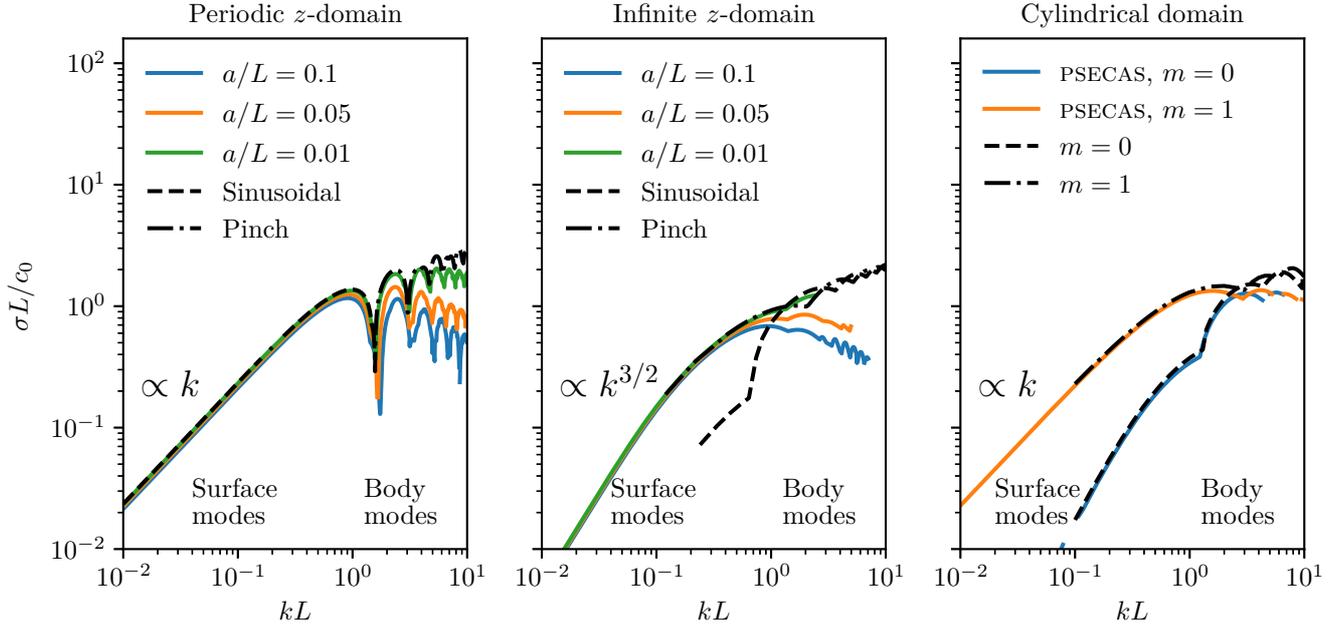}
\caption{ Dispersion plots for the KHI in the three different setups with
  $V/c_0=2.5$ and $\delta=1$. We compare results from the analytic
  dispersion relations derived for discontinuous density and velocity profiles
  with growth rates found using \textsc{psecas} and smooth profiles. Generally,
  the results coincide at low wavenumbers (long wavelengths) where the smoothing
  is insignificant. At high wave numbers the growth rates diverge for the
  discontinuous profiles but are diminished when the profiles are
  smooth. The low wavenumber solutions are surface modes while the high
  wavenumber solutions are body modes. \emph{Left and middle panel:} we compare
  the growth rates for the periodic and the infinite $z$-domain. We observe that
  the scaling with $k$ differs at low wavenumbers with
  $\sigma\propto k$ for the periodic $z-$domain and
  $\sigma\propto k^{3/2}$ for the infinite domain. The growth
  rates at various smoothing lengths converge to the analytic theory as the
  smoothing length is decreased with a slower convergence at high $k$.
  \emph{Right panel:} the growth rates for azimuthal modes with $m=0$ and $m=1$
  in cylindrical geometry. The \textsc{psecas} solutions have little
  smoothing ($a=0.01L$) and consequently agree with the analytic dispersion
  relation except at high $k$.  }
\label{fig:dispersion_plot}
\end{figure*}

For the planar sheet, i.e., a single interface at which the velocity changes
direction (see left panel of Fig.~\ref{fig:khi_setups}), the KHI is stabilized
when the flow velocity is sufficiently high \citep{Landau1944,Mandelker2016}.
This occurs when the flow velocity (see eq. 22 in \citealt{Mandelker2016})
\be
    \f{V}{c_0} > \f{1}{2}\sqrt{\f{\gamma}{1+\delta}}
    \left[1 + (1 + \delta)^{-1/3}\right]^{3/2}  \approx 1.1 \ ,
    \label{eq:M_crit}
\en
for $\delta = 1$ and $\gamma=5/3$.

For the planar slab (see the right panel of Fig.~\ref{fig:khi_setups}) discussed
in Section~\ref{sec:simulations} a different type of KHI can however arise at
high flow velocities. These are interchangeably called reflective modes
\citep{Payne1985,Hardee1988} or body modes \citep{Mandelker2016}.  Reflective
modes because the instability consists of soundwaves that are reflected between
the interfaces of the slab and body modes because it highlights that the KHI
disturbances occur throughout the slab and not just at its surfaces.  In this
section we analyze the body modes in the periodic Cartesian domain, the
Cartesian infinite domain and in cylindrical geometry. Examples of the pressure
perturbation for body modes with $V/c_0 = 2.5$ and $\delta=1$ in the
various setups are shown in Fig.~\ref{fig:body_modes}. In these figures the
density and the direction of the velocity changes at the interfaces indicated
with dotted black lines.  This change is smooth on a scale $a/L=0.05$ and the
solutions have been obtained with \textsc{psecas}.

In the following subsections we provide details for these calculations and
consider solutions where we decrease the smoothing parameter, $a$, thus
imitating a discontinuous profile. These solutions are compared with the
analytic theory for the discontinuous profiles for the slab in an infinite
Cartesian domain and the cylindrical slab presented in
\citep{Mandelker2016}. For the periodic Cartesian domain used in
Section~\ref{sec:simulations} and by \citet{Lecoanet2016}, we derive the
analytic theory for discontinuous profiles by following the procedure outlined
in \citet{Mandelker2016} while taking the different boundary conditions into
account. We obtain a dispersion relation, which yields growth rates that differ
from those found on the infinite domain.

A common element for the analysis in the following sections is the generalized
wavenumber
\be
    q = k \sqrt{1 - \f{(\omega - k \varv)^2}{k^2 \gamma \cs^2}} \ ,
    \label{eq:q-wave}
\en
where $k=k_x$ in Cartesian geometry and $k=k_z$ in cylindrical geometry.
The generalized wave number depends on velocity and density and has different
values in the slab (or stream) and in the background. We refer to the 
background with 0 subscripts, and to the slab with a subscript  $s$, e.g.,
$q_0$ and $q_s$.

\subsection{Planar shear on a Cartesian, periodic domain}
\label{sec:body_per}

We consider the planar slab used for the simulation H$\delta$ in
Section~\ref{sec:simulations}, i.e., $\delta=1$ and an equilibrium given by
Equations~\eqref{eq:v-equi} and \eqref{eq:rho-equi} on a domain which is periodic in both $x$ and $z$.
In order to study body modes, we increase the flow speed to $V/c_0=2.5$
such that Equation~\eqref{eq:M_crit} is fulfilled and
calculate the linear solution for $k=2\pi/L$. The resulting pressure
perturbation is shown in the top left
panel of Fig.~\ref{fig:body_modes}. In this figure the equilibrium velocity and
density change smoothly at $z/L=0.5$ and $z/L=1.5$. The pressure perturbation
resulting from the KHI is not confined to these surfaces but instead extends
throughout the domain. The
perturbations are thus present both in the central slab (with $1/2<z/L<3/2$)
and in the slab which, due to the periodicity in the $z$-direction, is the
union of the regions with $z/L<1/2$ and $z/L>3/2$. The pressure perturbation
is highest in the latter region where the density is lower. This phenomenon is
also seen for KHI perturbations of surface modes which penetrate deeper into
the fluid with lower density (see e.g., Fig. 2 in \citealt{Mandelker2016}).

The structure of the mode is qualitatively different from the corresponding
mode on the infinite domain (see Fig.~\ref{fig:body_modes}). Here the pressure perturbation is
confined to the
central slab and decays abruptly outside of it. The key difference is that the
periodic domain consists of two slabs that are connected with each other on
both sides. We have seen in Section~\ref{sec:simulations} that this setup can be very
useful for code testing purposes. Whether this setup can be used to model the
KHI in a real physical application is however not obvious. In the following
we discuss the difference between the two setups by
calculating growth rates for the KHI in the two different setups using both
analytic theory for discontinuous profiles and \textsc{psecas} for smooth
profiles.

We begin by deriving the analytic
dispersion relation for the two connected slabs on a periodic domain when the
profiles are discontinuous. The derivation closely follows the analysis of the
slab on the infinite domain presented in \citet{Mandelker2016}.
We consider three domains\footnote{In order to highlight the symmetry of the
pressure solution, we derive the analytic dispersion
relation in a coordinate system with $z \in [-L,\, L]$ where the shear
interfaces are located at $z=\pm L/2$. This simply corresponds to a
translation of the setup considered in Section~\ref{sec:simulations}
(by a distance $-L$) and does not modify the physics.
} where domain 1 has $z>L/2$,
domain 2 has
$|z|<L/2$ and domain 3 has $z<-L/2$. By combining Equation~\eqref{eq:rho-lin}
and Equation~\eqref{eq:ent-lin} we obtain an equation for the pressure
\be
-i\left(\omega - k\varv\right)\f{\delta p}{p} =
   - i k \gamma \dvx
- \gamma \pder{\dvz}{z} \ ,
\label{eq:lin-dp-hydro}
\en
where $\dvx$ and $\dvz$ are given by
\begin{align}
-i\left(\omega - k\varv\right)\dvx &=
- \pder{\varv}{z} \dvz
% pressure
- i k \f{\delta p}{\rho} \ ,
\label{eq:lin-dvx-hydro}
\end{align}
and
\begin{align}
-i\left(\omega - k\varv\right) \dvz &=
% pressure
- \f{1}{\rho}\pder{\delta p}{z} \ .
\label{eq:lin-dvz-hydro}
\end{align}
These are simply simplified forms of
Equations~\eqref{eq:mom-x-lin} and \eqref{eq:mom-z-lin} where the background magnetic field,
the pressure gradient etc. have been set to zero.
We combine Equations~\eqref{eq:lin-dp-hydro} to \eqref{eq:lin-dvz-hydro} and find a
second order, ordinary differential equation (ODE) for the perturbed pressure
\be
\pdder{\delta p}{z} +
\left(\f{2 k}{\omega - k \varv} \pder{\varv}{z} - \f{1}
{\rho}\pder{\rho}{z}\right) \pder{\delta p}{z}
- q^2 \delta p = 0 \ ,
\label{eq:pres-ode}
\en which we solve analytically for piece-wise constant velocity and density
profiles. Assuming discontinuous initial conditions, inside each of the three
domains the density and velocity is constant and the terms in
Equation~\eqref{eq:pres-ode} which are proportional to $\partial \delta
p/\partial z$ are thus zero.  The solution of Equation~\eqref{eq:pres-ode} is
therefore simply $\delta p (z)/p = a e^{-q z} + b e^{qz}$ where $q$ is given by
Equation~\eqref{eq:q-wave}.  The constants $a$ and $b$ are deduced by requiring
that \emph{i)} the solution is continuous across the domain interfaces at $z=\pm
L/2$ and the interface at $z=\pm L$ and \emph{ii)} the fluid displacement, $h$,
is continuous across the interfaces.  The latter requirement is known as the
Landau condition (see e.g., \citealt{Mandelker2016}).

In order to derive the Landau condition,  we
consider the linearized equation
for the interface displacement, $h$,
\be
   \der{h}{t} = \delta \varv_z = -i(\omega - k \varv)  h
\en
which combined with the equation for $\dvz$ yields
\be
 h = - \f{1}{\rho (\omega - k \varv)^2}\pder{\delta p}{z}
 \label{eq:landau-condition}
\en
Applying the Landau condition at $z=\pm L$ as well as continuity of the
solution, we find that the pressure profile is given by
\begin{alignat}{4}
    \delta p(z)/p &= A\, \mathcal{S}(-q_0 [z-L]),
    \hspace{40pt}&z &> &L/2 ,\\
    \delta p(z)/p &= A\, \f{\mathcal{S}(q_0 L/2)}{\mathcal{S}(q_s L/2)}
    \mathcal{S}(q_s z),
    &|z| &< &L/2 ,\\
    \delta p(z)/p &= A\, \mathcal{S}(-q_0 [z+L]),
     &z &< -&L/2 ,
\end{alignat}
where $\mathcal{S}(z) = \cosh(z)$ for symmetric pinch modes,
$\mathcal{S}(z) = \sinh(z)$ for antisymmetric sinusoidal modes and $A$ is an
overall amplitude. An illustration of the difference between sinusoidal and
pinch modes can be found in Fig. 3 in \citealt{Mandelker2016}.

Applying the Landau condition at either $z=-L/2$ or $z=L/2$ then
yields the dispersion relation for the two connected slabs in a periodic
domain, given by
\be
    \left(\f{\omega-k \varv_s}{\omega-k \varv_0}\right)^2 =
    - \f{\rho_0}{\rho_s}\f{q_s}{q_0} \f{\mathcal{T} (q_s L/2)}{\mathcal{T} (q_0 L/2)} \ .
    \label{eq:disp-body-per}
\en
with $\mathcal{T}(z)=\tanh(z)$ for the pinch modes and
$\mathcal{T}(z)=1/\tanh(z)$ for the sinusoidal modes.

We have found the maximum growth rate as a function of wavenumber by numerically
solving Equation~\eqref{eq:disp-body-per} for both pinch and sinusoidal
modes. The results, presented in the left panel of
Fig.~\ref{fig:dispersion_plot} (black lines), show that the growth rate is
highest for the sinusoidal modes at low wavenumbers and for the pinch modes at
high wavenumbers.

Equation~\eqref{eq:disp-body-per} can be solved analytically for $\omega$ in the
incompressible limit, where $q_0=q_s=k$ and the terms with $\mathcal{T}$
cancel. The analytical result thus obtained is given by Equation~\eqref{eq:chandra-khi},
which was originally derived for the incompressible, planar sheet. The long
wavelength $\sigma \propto k$ scaling of the growth rate is also found for
the $m=0$ solution for the cylinder (\citealt{Mandelker2016}
and our Section~\ref{sec:body_cyl}) but the slab on the infinite domain has
$\sigma \propto k^{3/2}$ \citep{Mandelker2016}. The chosen geometry can thus
qualitatively change the pressure profile and the scaling of the growth rates
with wavenumber.

We can also calculate the maximal growth rate as function of
wavenumber using \textsc{psecas}.
The result of such calculations with various smoothing
lengths, $a$, are shown in the left panel of
Fig.~\ref{fig:dispersion_plot}. At long
wavelengths (small $k$), the \textsc{psecas} results and the sinusoidal mode
solution
of Equation~\eqref{eq:disp-body-per} coincide with the incompressible
limit of Equation~\eqref{eq:disp-body-per}, i.e., Equation~\eqref{eq:chandra-khi}. At short
wavelengths where the solutions are body modes, smoothing of the profiles
creates a difference between the solutions obtained with \textsc{psecas} and
solutions to Equation~\eqref{eq:disp-body-per}. When the smoothing length is decreased
and the smooth profile tends towards the discontinuous limit, this difference
is reduced. Furthermore, the locations of peaks in the growth rate, are only
weakly modified by the value of the smoothing coefficient. The underlying
reason is that the peak locations are determined by a resonance condition
which depends on the width of the slab
\citep{Payne1985,Hardee1988,Mandelker2016}.

\begin{figure*}
\includegraphics[trim = 0 15 0 0]{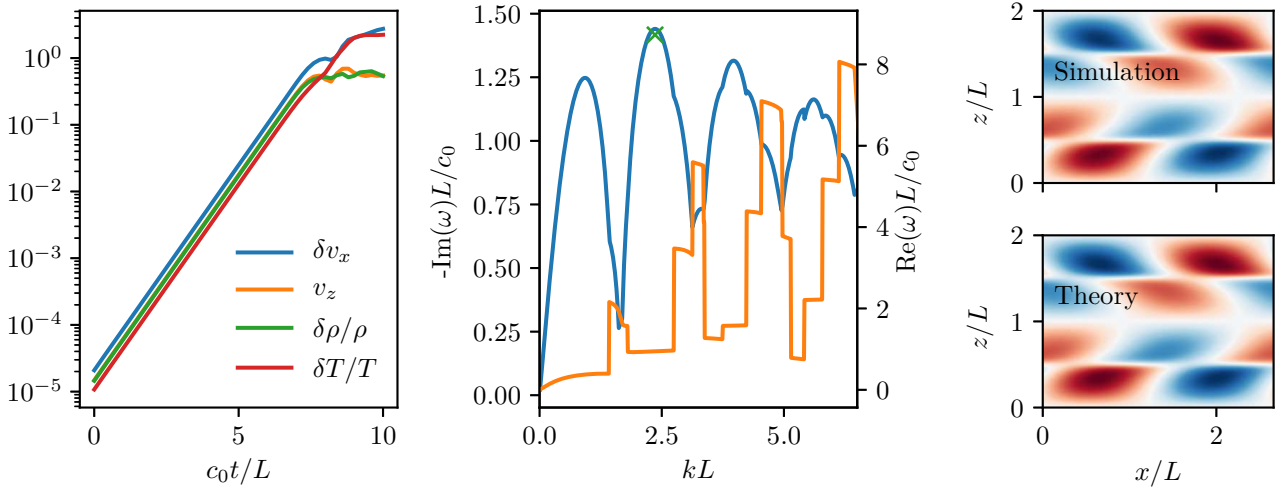}
\caption{Simulation with $V/c_0 = 2.5$ and $\delta = 1$. At these
parameters, the fastest growing mode is a body mode, i.e., the mode structure
extends far beyound the shear surfaces. \emph{Left:} Evolution of
perturbation amplitude in the \textsc{Athena} simulation. \emph{Middle:}
Growth rate (blue line) and real frequency (orange line) as a function
of wavenumber with the growth rate from the \textsc{Athena} simulation
indicated with a green cross. \emph{Right:} The pressure perturbation in the
simulation and evaluated using Equation~\eqref{eq:lin_solution} at $c_0 t/L = 5$.}
\label{fig:body_mode}
\end{figure*}

For computer simulations of body modes, the smooth profiles have the
attractive property that the growth rates at short wavelengths are inhibited.
We consider a simulation of the
fastest growing body mode when $a=0.05L$ in Fig.~\ref{fig:body_mode}. We show the
exponential growth of perturbations in the left panel of Fig.~\ref{fig:body_mode}
and the fitted growth rate in the middle panel of Fig.~\ref{fig:body_mode}.
We find that the growth rate obtained from the simulation has a relative error of almost 2 \%,
significantly higher than the error found at lower flow velocities (i.e. the
H$\delta$ simulation had $V/c_0=1$ compared to $V/c_0=2.5$ used for the
body mode simulation). The maximum growth rate occurs at $k L=
2.3629555$ where $\omega L/c_0 = 0.949814+1.440521i$.

\subsection{Planar shear on Cartesian, infinite domain}
\label{sec:body_inf}

We consider a slab located at $|z|<L/2$ on an infinite domain with $z\in
[-\infty, \infty]$. We can study this setup with \textsc{psecas} by using the
rational Chebyshev TB grid which models the infinite domain \citep{Boyd1987}.  A
smooth velocity and density profile is described by Equations~\eqref{eq:v-equi}
and \eqref{eq:rho-equi} with $z_1=-L/2$ and $z_2=L/2$. The resulting
perturbed pressure
profile for a body mode with $k=2\pi/L$ and $a/L=0.05$ is shown in
Fig.~\ref{fig:body_modes}. The body mode consists of sound waves inside the slab
with perturbations that decay outside of it. Here, the pressure perturbation differs
qualitatively from the corresponding perturbation on the periodic $z$-domain.

For a discontinuous profile, \citet{Mandelker2016} solved
Equation~\eqref{eq:pres-ode} and applied the
boundary conditions that the pressure perturbation goes to zero at
$z\rightarrow \pm \infty$. By applying the Landau condition at either
$z=-L/2$ or $z=L/2$, this yields the dispersion relation \citep{Mandelker2016}
\be
    \left(\f{\omega-k \varv_s}{\omega-k \varv_0}\right)^2 =
    - \f{\rho_0}{\rho_s}\f{q_s}{q_0} \mathcal{T} (q_s L/2) \ .
    \label{eq:disp-body-inf}
\en
with $\mathcal{T}(z)=\tanh(z)$ for the pinch modes and
$\mathcal{T}(z)=1/\tanh(z)$ for the sinusoidal modes. A detailed analysis of
this dispersion relation is presented in \citet{Mandelker2016}, in the following
we numerically solve Equation~\eqref{eq:disp-body-inf} in order to compare with
solutions obtained with \textsc{psecas} and a smooth profile.  This comparison
is shown in the middle panel of Fig.~\ref{fig:dispersion_plot} using three
different smoothing lengths ($a/L=0.1$, $0.05$, and $0.01$).

At long wavelengths (where the solutions are surface modes), the
\textsc{psecas} results and the pinch mode solution of Equation~\eqref{eq:disp-body-inf}
both coincide with the incompressible limit of Equation~\eqref{eq:disp-body-inf} derived in
\citet{Mandelker2018}. At high
wavelengths where the solutions are body modes, the smooth profile
\textsc{psecas} solutions and the solutions to Equation~\eqref{eq:disp-body-inf} differ.
This is again because a smooth profile lowers the growth rates.
As for the periodic slab, this difference
is reduced as the smoothing length is decreased towards the discontinuous
limit.

Equation~\eqref{eq:disp-body-inf} has effective growth rates that diverge as
$\sigma \propto \ln(k L)$ which makes convergence of simulations for
the slab on the infinite
domain unattainable with a discontinuous profile \citep{Mandelker2016}.
For a sufficiently smooth profile, this divergence of the growth
rate is removed and the growth rate is suppressed at high wave numbers.

\subsection{Cylindrical KHI}
\label{sec:body_cyl}

We are interested in understanding the stability properties of a cylindrical
stream of gas which moves at velocity $V$ through an ambient medium moving
in the opposite direction with speed $-V$. Due to Galilean
invariance of the equations this is equivalent to a stream of gas moving at $2V$
through a stationary medium. The setup is illustrated in
Fig.~\ref{fig:khi_setups}. The stability properties of this system are important
for understanding whether cold streams of gas can feed galaxies at high redshift
or whether they are interrupted and heated by the KHI before reaching the
galaxies \citep{Mandelker2016,Padnos2018,Mandelker2018}.  We consider a
cylindrical stream with diameter $L$ (radius $R=L/2$) and define a standard
cylindrical coordinate system with unit vectors $\er$, $\ephi$ and $\ez$. We
take the background to only depend on the radial coordinate, $r$, with $\rho(r)$
and the velocity directed along $\ez$, i.e., $\vec{\varv}= \varv(r)\ez$.

We assume perturbations from this equilibrium of the form
\be
    f_{k,\,m}(r) e^{-i\omega t + i m\phi +i k z},
\en
which results in linearized hydrodynamical equations given by
\begin{align}
    -i\omega \delta \varv_r &=
    - ik \varv \delta \varv_r
    - \cs^2 \pder{}{r}\f{\delta p}{p} \ ,
    \label{eq:lin-dvr-cyl} \\
    -i\omega \delta \varv_\phi &=
    - i k \varv \delta \varv_\phi
    - im \f{\cs^2}{r}\f{\delta p}{p} \ , \\
    -i\omega \delta \varv_z &=
    -  i k \varv \delta \varv_z
    -   \pder{\varv}{r}\delta \varv_r
    -  i k \cs^2 \f{\delta p}{p}  \ ,
\end{align}
and
\begin{align}
    -i\omega \f{\delta p}{p} &=
    - i k \varv \f{\delta p}{p}
    - \gamma \pder{\delta \varv_r}{r}
    - \gamma \f{\delta \varv_r}{r}
    - \gamma \f{i m}{r}\delta \varv_\phi
    - \gamma i k \delta \varv_z
    \label{eq:lin-dp-cyl}
\end{align}
in cylindrical coordinates.
For the calculations with \textsc{psecas}, we take the background to be
\begin{align}
    \varv(r) &= V \tanh\left(\f{r-R}{a}\right) \ , \\
    \rho(r) &= \rho_0 + \rho_0 \f{\delta}{2}\left[ 1 - \tanh\left(\f{r-R}
    {a}\right)\right] \ ,
\end{align}
where $a$ is the smoothing parameter. In the limit $a\rightarrow0$ this
corresponds to a discontinuous profile with density $\rho_0(1+\delta)$ and
velocity $V$ inside the stream and $(\rho, \varv) = (\rho_0,
-V)$ outside the stream.

Equations~\eqref{eq:lin-dvr-cyl} to \eqref{eq:lin-dp-cyl} are defined on $r \in
[0, \infty]$ with a coordinate singularity at $r=0$. We use the rational
Chebyshev TL polynomials to discretize the problem on
this
semi-infinite domain \citep{Boyd1987a}. Due to the coordinate singularity, this
polynomial basis automatically applies boundary conditions that
ensures that the perturbations are finite when $r\rightarrow 0$ and $r
\rightarrow \infty$ (so-called behavioral boundary conditions, see
\citealt{Boyd1987a,Boyd}).
We take  $k_z=2\pi/L$ and $m=1$ and calculate the resulting perturbed pressure
profile which we show\footnote{Here we are simply showing a slice
with constant $z$ but the obtained linear solution for
the cylinder is three-dimensional.} in Fig.~\ref{fig:body_modes}. This mode is a
body
mode with a pressure perturbation that
extends throughout the interior of the cylindrical stream.

We are again interested in comparing results from \textsc{psecas} using
smooth profiles with analytic theory for discontinuous profiles.
As for the Cartesian setup, the linearized equations, i.e.,
Equations~\eqref{eq:lin-dvr-cyl} to \eqref{eq:lin-dp-cyl}, can be combined
to obtain an ODE for the pressure perturbation \citep{Mandelker2016}
\be
    \pdder{\delta p}{r}
    +
    \left(\f{2k}{\omega - k \varv}\pder{\varv}{r} - \f{1}{\rho}\pder{\rho}{r}\right)
    \pder{\delta p}{r}
     - \left(q^2  + \f{m^2}{r^2}\right)
    \delta p  = 0 \ .
    \label{eq:pressure-ode-cyl}
\en
Away from the interface where the terms proportional to $\partial \delta p
/\partial r$ are zero, Equation~\eqref{eq:pressure-ode-cyl} can be written as a modified
Bessel
equation. The solutions are therefore $\delta p/p = a \mathcal{I}_m(q r) + b
\mathcal{K}_m(qr)$ where $\mathcal{I}_m$ ($\mathcal{K}_m$) is the modified
Bessel function of the first (second) kind. Since the $
\mathcal{I}_m\rightarrow \infty$ ($\mathcal{K}_m\rightarrow \infty$) as
$r\rightarrow\infty$ ($r\rightarrow 0$) the solution has $b=0$ ($a=0$) inside
(outside) the cylinder \citep{Mandelker2016}.
Applying the Landau condition at $r=R$ yields the dispersion relation
\citep{Mandelker2016}
\be
\left(\f{\omega - k \varv_s}{\omega - k \varv_0}\right)^2
=
\f{\rho_0}{\rho_s}\f{q_s}{q_0}
\f{\mathcal{I}'_m(q_0 R)}{\mathcal{I}_m(q_0 R)}
\f{\mathcal{K}_m(q_s R)}{\mathcal{K}'_m(q_s R)}
\label{eq:disp-cyl} \ ,
\en
which was derived and analyzed in detail by \citet{Mandelker2016}.

The comparison between solutions of Equation~\eqref{eq:disp-cyl} for the azimuthal
wavenumbers $m=0$ and $m=1$
with the solutions obtained with \textsc{psecas} with a very short
smoothing length ($a=0.01L$) are presented in Fig.~\ref{fig:dispersion_plot}.
At long wavelengths, all solutions agree with the scalings derived in
\citet{Mandelker2016}.

\section{Discussion}
\label{sec:discussion}

\subsection{Code testing and verification}

Tests of software used for modeling physical systems are
generally classified as either verification or validation 
(see \citealt{Oberkampf2002} for a review). 
In brief, code verification consists of
ensuring that the code accurately solves the mathematical model (i.e., 
Equations~\eqref{eq:rho} to \eqref{eq:ent} in our case) while
validation 
assesses to which degree simulations agree with physical reality. In
engineering, validation is achieved by comparing simulations with 
dedicated validation experiments \citep{Oberkampf2002}. In computational
astrophysics, validation is unfortunately very rarely possible using
experiments and is instead assessed by comparing simulations to
observations.
Verification can be considered a preliminary of
validation, and is therefore essential for astrophysical codes.
Verification is normally done by comparing computer simulations with known
analytical solutions or highly accurate numerical solutions.
Analytical solutions to the nonlinear fluid equations have only been obtained
in a very limited number of cases and many verification
tests therefore compare simulations with analytical solutions of the 
\emph{linearized} fluid equations.

Even finding an analytic solution to the set of linearized
equations can however be problematic in some cases, e.g., when there is a
non-trivial variation of the background equilibrium. This is the case for the
KHI where analytical solutions have not been found for the smooth initial
conditions that are required for convergence of simulations 
\citep{Robertson2010,McNally2012,Lecoanet2016}.
The lack of an analytical reference solution for the KHI motivated the studies
of \citet{McNally2012} and \citet{Lecoanet2016} who obtained accurate, 
numerical reference solutions in 2D by performing high resolution
simulations of the KHI. Obtaining such reference simulations is very
computationally intensive because high resolution is required in the
two spatial dimensions and time.
The reference solutions were made public by \citet{McNally2012}
and \citet{Lecoanet2016} and are indispensable for verification of the
nonlinear
stage of the KHI during which vortex rolls form and secondary instabilities
take place.

A linear reference solution would however be sufficient for verification of
the linear regime.
While such a linear reference solution has not been found analytically
for the KHI with smooth initial conditions, it \emph{is} possible to obtain 
an extremely accurate numerical solution using pseudo-spectral methods. In
fact, the numerical linear solution can be found by discretization
of only the direction perpendicular to the background flow
while the direction parallel to the flow and the time evolution is treated
analytically.
The pseudo-spectral solution of the linearized equations therefore retains
some of the
useful properties of analytical solutions, i.e., that
they can be evaluated to very high precision and that solutions with different
physical parameters can be obtained rather quickly. For instabilities,
the pseudo-spectral linear theory also makes it possible to determine the
growth rate as a function of wavelength and the structure of the fastest
growing mode. This knowledge provides important physical insight and can
furthermore aid in designing nonlinear simulations. 
In particular, the growth rate versus wavelength determines what size
the computational domain should be in order to resolve the full spectrum of
the instability (e.g. \citealt{Shalaby2017}).

Using the MHD code \textsc{Athena}, we give an example of
how
pseudo-spectral linear theory
can be used for inexpensive code verification of problems where
analytic solutions are not available.
We present simulations of the KHI with the smooth initial conditions used for
convergence studies in
\citet{Lecoanet2016} and additionally treat the MHD and Braginskii-MHD
versions of the instability.
We consider simulations of four different versions of the KHI (with
parameters listed in
Table~\ref{tab:khi_table})
in which we seed the instability with the linear solution for the fastest
growing eigenmode. We find excellent agreement between linear theory and
simulations during the linear stage of the instability and this agreement is
measured both in terms of the eigenmode structure and the growth rate. We also
perform a convergence study and show that the relative error in both the
eigenmode and the growth rate converge at second order.
Importantly, we make material available online which can be used
by readers to initialize the KHI with the linear solution.
Alternatively, Table~\ref{tab:khi_table} can be used to confirm that the KHI
in tests
initialized with random perturbations grow at the maximum growth rate.

The \textsc{Athena} simulations presented in this paper were performed without
MPI and still ran in less than a core hour in total.\footnote{The $N_x=512$
  simulation with $\delta=1$ ran in 23 core minutes and the $N_x=256$ simulation
  with $\delta=0$ ran in less than 2 core minutes on a 2.5 GHz
  Intel Core i7 processor.} For comparison, the most expensive simulations presented in
\citet{Lecoanet2016} took roughly $10^6$ core hours. In terms of computing time,
it is therefore economical to do a convergence study of the linear regime before
proceeding with a nonlinear convergence study.  The linear tests presented in
this paper can thus be useful for frequent, computationally inexpensive
regression-tests of a code and two of the tests (H$\delta$ and M) have
in fact already been included
in the test suite of \textsc{Arepo}
\citep{Springel2010,Pakmor2011,Pakmor2013,Pakmor2016a}. Furthermore, one
of the tests (M$\nu$) has proven very useful during the implementation of
Braginskii viscosity in \textsc{Arepo} (Berlok et al., in prep.).

\subsection{Astrophysical implications}

\subsubsection{Cold fronts in galaxy clusters}
Sloshing cold fronts can arise when a cool-core galaxy cluster is subject to a
minor
merger. This leads to a disturbance in the ICM which exhibits a sloshing
motion in the perturbed gravitational potential with a resulting cold front 
(see \citealt{Markevitch2007,ZuHone2016} for reviews). Cold fronts are 
cold and dense on one side and hot and dilute on the
other side with a tangential velocity difference across the interface and
should therefore be highly susceptible to
the KHI. Observationally however, the KHI
is not as abundantly observed as one would expect by applying, e.g., Equation 
\eqref{eq:chandra-khi} to observed values for densities and velocities. 
Possible explanations for this observational evidence
for the suppression of the KHI include suppression by magnetic field tension 
(e.g. \citealt{Vikhlinin2001a,Dursi2008}), viscosity (e.g. 
\citealt{Roediger2013}), Braginskii viscosity 
(e.g. \citealt{Suzuki2013,Zuhone2015}) or a finite width of the cold front 
\citep{Churazov2004}. 

Most of these effects, except for isotropic viscosity, are captured by the
linear theory derived in Section~\ref{sec:equations_and_equilibria}.
We show in Figure~\ref{fig:growthrates} how Braginskii viscosity, magnetic
fields, a smooth transition in the velocity profile and a density contrast
all act to suppress the KHI. 
In the future, we intend to use Equations~\eqref{eq:rho-lin} to 
\eqref{eq:ent-lin} and
\textsc{psecas} to perform a detailed analysis of the stability
properties of cold fronts, taking into account variations in magnetic field
strength (due to magnetic draping), velocity, density and temperature across
the cold front and using observationally inferred parameters (Berlok and
Pfrommer, in prep.).

\subsubsection{Cold streams feeding high redshift galaxies}
Massive galaxies at redshifts $z=1-4$, with baryon mass $\sim10^{11}
M_\odot$
and dark matter halo virial mass $\sim10^{12} M_\odot$, have star formation
rates
which are incompatible with the slow cooling rate of the virialized gas
residing in the halo. Such massive galaxies are however located at nodes
of the cosmic web, and the observed star formation rate ($\sim 100\,M_\odot$
yr$^{-1}$) is similar to the
gas accretion found to occur along filaments in cosmological simulations. In
contrast to the low density, $10^6$ K-virialized gas in the halo, the
high density in the filaments allows the gas to efficiently cool to
$10^4$ K before reaching the massive galaxy. A possible explanation for the
observed star
formation rate is thus that several cold, dense streams are able to penetrate
through the hot, dilute gas and reach the central galaxy without being
disrupted. 

One potential disruption mechanism is the KHI, and a detailed understanding of
the KHI is therefore important for understanding whether and how cold streams of
gas can feed galaxies at high redshift \citep{Mandelker2018}.  Cosmological
simulations do not yet have the required spatial resolution to assess whether
the KHI is present in cold streams and thus cannot be used for directly
answering this question. As an alternative,
\cite{Mandelker2016,Padnos2018,Mandelker2018} performed a very detailed study of
the KHI using both analytical theory and simulations. Our results presented in
Section~\ref{sec:bodymodes}, agree with growth rates calculated in
\citet{Mandelker2016} for supersonic body modes in both planar slab and
cylindrical geometry (see Figure~\ref{fig:khi_setups} and
\ref{fig:body_modes}). Additionally, we are able to calculate the suppression of
short wavelength modes when there is a smooth transition in the velocity profile
between the cold stream and the ambient medium. By assuming a smoothing length,
$a$, we are able to calculate the maximal growth rate of body modes. This is in
contrast to the analytic theory where the maximal growth rate diverges with
wavenumber \citep{Mandelker2016}. Hence, our framework is able to account for
more realistic density and velocity profiles when estimating how quickly cold
streams will be disrupted by the KHI.

\section{Conclusions}
\label{sec:conclusions}

Smooth initial conditions are known to be a necessity for the convergence
of computer simulations of the KHI, even for the linear stage
\citep{Robertson2010,McNally2012}.
Analytic predictions for the evolution of a simulation with such smooth
initial conditions do however not exist, neither for the linear nor for the
nonlinear regime of the instability. In lieu of analytic results,
\citet{Lecoanet2016} has provided converged numerical results for the
nonlinear regime of the hydrodynamic KHI as it occurs in a doubly periodic
domain. Their simulations included explicit viscosity, heat conductivity and
diffusion of a passive scalar and required very high grid resolution for
convergence. As a supplement to convergence studies of the nonlinear
regime, code tests of the linear stage are often used because they are
computationally cheaper. Tests limited to the linear regime make it feasible
to perform isolated testing of various physics modules and to
explore a large parameter space.
Such computer simulations are often compared with analytic results for
discontinuous initial conditions because the analytical linear theory does not
exist for smooth profiles.
Quantitative comparisons between linear theory and
simulations however require that the linear theory is derived with the
same assumptions as the simulations. This includes boundary conditions,
equation of state and (smoothness of) initial conditions.

With this problem in mind, we have derived the linear theory for the KHI with
smooth initial conditions in both Cartesian and cylindrical geometry
(illustrated in Fig.~\ref{fig:khi_setups}). The linear theory in Cartesian
geometry combines and extends some of the results of previous work on the
compressible hydrodynamic KHI \citep{Blumen1970}, the MHD version of the KHI
\citep{Miura1982} and the KHI with Braginskii viscosity \citep{Suzuki2013}.

The simulations discussed in Section~\ref{sec:simulations} (and the ones presented in
\citealt{Lecoanet2016}) used a subsonic flow velocity where the KHI appears
as \emph{surface} modes, i.e., the disturbances to the flow are localized at
the shear interfaces. At supersonic flow velocities, the KHI can manifest
itself
as \emph{body} modes that have disturbances with much larger extent
\citep{Payne1985,Hardee1988,Mandelker2016}.
Specializing to the equations of inviscid hydrodynamics, we analyze body modes
in both Cartesian and cylindrical geometries. We compare our results found
with initial conditions that are smooth on a scale, $2a$, with
analytic theory derived for discontinuous density and velocity profiles.
As expected, we find increasing agreement with the discontinuous, analytic
theory as the value of $a$ is decreased. We also find that the introduction of
a smoothing length inhibits the growth of the KHI body modes at high
wavenumbers and conclude that a smoothing length will allow simulations of
both subsonic and supersonic flows to converge.

The analytic theory for the Cartesian slab when the domain is periodic in both
directions (the two connected slabs used in \citealt{Lecoanet2016}), is
derived and solved (albeit numerically) for the first time.
Interestingly, we find that the growth rates scale with wavenumber,
$\sigma \propto k$, at long wavelengths. This scaling differs from the
result
found by \citet{Mandelker2016} on the infinite domain (here $\sigma\propto
k^{3/2}$) but is identical to the result found for the planar
sheet and the cylindrical stream with azimuthal symmetry ($m=0$,
\citealt{Mandelker2016}). This
result, along with the qualitatively different eigenmode for the body mode on
the two connected slabs (Fig.~\ref{fig:body_modes}),
might call into question how and if lessons learned from
the doubly periodic slab can be applied to real physical systems. Despite
this concern, this setup is nevertheless immensely useful for code testing
purposes.

The solution of the eigenvalue problem that arises from
linearizing the dynamical equations is non-trivial and cannot be solved
analytically for smooth initial conditions. This is especially so when a
variety of boundary conditions, geometries and physical effects beyond
hydrodynamics are of interest. Luckily, the kind of eigenvalue problems that
arise in astrophysical fluid dynamics can often be tackled numerically with
pseudo-spectral methods. As an aid for such calculations, \textsc{psecas}
(\textbf{P}seudo-\textbf{S}pectral \textbf{E}igenvalue \textbf{C}alculator
with an
\textbf{A}utomated \textbf{S}olver) automates many of the
steps involved and is freely available online.

\section*{Acknowledgments}
T.B. would like to thank Gopakumar Mohandas for many discussions on
pseudo-spectral methods and for implementing the Legendre grid in
\textsc{psecas}. We thank the referee, Colin McNally, for an
insightful report which helped us improve the manuscript and  Nir
Mandelker for enlightening discussions. We
are grateful to the authors of \textsc{Athena} 
\citep{stone_athena:_2008},
\textsc{Numpy} \citep{Oliphant2007}, \textsc{Scipy} \citep{scipy},
\textsc{Matplotlib} \citep{Hunter2007} and \textsc{mpi4py} \citep{Dalcin2008}
for making their software freely available.  T.B. and C.P.  acknowledge support
by the European Research Council under ERC-CoG grant CRAGSMAN-646955.

\bibliographystyle{mnras}
\bibliography{references}

% Don't change these lines
\bsp	% typesetting comment
\label{lastpage}
\end{document}